\documentclass[12pt]{article}
\usepackage{amsmath}
\usepackage{amssymb}
\usepackage{physics}
\usepackage{siunitx}
\usepackage{newtxtext,newtxmath}
\usepackage{graphicx}
\usepackage{mathtools}
\usepackage[letterpaper,margin=1in]{geometry}
\renewenvironment{abstract}
     {\quotation}
     {\endquotation}
\date{}
\makeatletter
\renewcommand{\fnum@figure}{\textbf{Figure \thefigure}}
\renewcommand{\fnum@table}{\textbf{Table \thetable}}
\makeatother
\usepackage{url}

\usepackage{setspace}
\usepackage{caption}
\DeclareCaptionFont{singlespace}{\setstretch{1.2}}
\def\scititle{Probing critical phenomena in open quantum systems using atom arrays}
\title{\bfseries \boldmath \scititle}
\author{
Fang Fang$^{1,2,3\dagger}$,
Kenneth Wang$^{1,2,3\dagger}$,
Vincent S. Liu$^{2\dagger}$,
Yu Wang$^{1,2,3\dagger}$,\\
Ryan Cimmino$^{1,2,3}$,
Julia Wei$^{2}$,
Marcus Bintz$^{2}$,
Avery Parr$^{2}$,
Jack Kemp$^{2}$,\\
Kang-Kuen Ni$^{1,2,3\ast}$,
Norman Y. Yao$^{2,1,3\ast}$
\\[6pt]
\small $^{1}$Department of Chemistry and Chemical Biology, Harvard University, Cambridge, MA 02138, USA\\
\small $^{2}$Department of Physics, Harvard University, Cambridge, Massachusetts 02138, USA\\
\small $^{3}$Harvard-MIT Center for Ultracold Atoms, Cambridge, Massachusetts 02138, USA\\[6pt]
\small $^\ast$Corresponding authors: ni@chemistry.harvard.edu, nyao@fas.harvard.edu\\
\small $^\dagger$These authors contributed equally.
}
\date{}
\begin{document}
\maketitle 
\begin{abstract} \bfseries \boldmath
At continuous phase transitions, quantum many-body systems exhibit complex, emergent behavior. Most strikingly, at a quantum critical point, correlations decay as a power law, with exponents determined by a set of universal scaling dimensions. Experimentally probing such power-law correlations is extremely challenging, owing to the interplay between decoherence, the vanishing energy gap, and boundary effects. Here, we employ a Rydberg quantum simulator to adiabatically prepare critical ground states of both a one-dimensional ring and a two-dimensional square lattice. By accounting for and tuning the openness of our quantum system, which is well-captured by a single phenomenological length scale, we directly observe power-law correlations and extract the corresponding scaling dimensions. Our work complements recent studies of quantum criticality using the Kibble-Zurek mechanism and digital quantum circuits.
\end{abstract}

Strongly-interacting many-body systems are a fertile ground for complex emergent behavior. This is perhaps most apparent at continuous quantum phase transitions, where fluctuations of competing orders extend over all length scales~\cite{Hertz1976, Coleman:2005, sachdev_2011}. In such quantum critical states, the intricate structure of correlations and entanglement exhibits striking universal features independent of the underlying microscopic details. In particular, all long-distance, low-energy behavior is expected to be governed by a small set of universal quantities, known as scaling dimensions~\cite{cardy1996scaling}. Determining these scaling dimensions remains a fundamental focus for both theoretical and experimental studies of quantum criticality~\cite{Poland:2019, Zhu:2023, Scheie:2023}.

The most direct signature of quantum criticality is the universal, power-law decay of correlations in real space; the exponents of this decay are directly proportional to the scaling dimensions. Despite conceptual simplicity, the direct observation of power-law decays remains challenging, even in quantum simulators with microscopic resolution and full access to local observables~\cite{King2018, Zhupnas20, mi2023stable, haghshenas2023probing}. There are three main difficulties. First, the direct preparation of critical ground states is difficult because of the vanishing of the energy gap \cite{Albash18}. Second, the long-range entanglement inherent to quantum critical states is inhibited by the presence of decoherence in noisy-intermediate-scale-quantum simulators~\cite{Boixo2014, King2022, Joshi2023,Azses:2023}. Finally, boundary effects in finite-size systems complicate the interpretation and extraction of universal physics from the bulk.

In this work, we address each of these challenges, and experimentally observe critical, real-space correlations using a programmable Rydberg simulator based on a cesium ($^{133}$Cs) atom array. Our main results are as follows. First, by carefully optimizing time-dependent ramp profiles, we adiabatically prepare critical ground states in the Ising universality class, in both one and two dimensions (Fig.~\ref{fig:Exp_scheme}). For the 1D case, we utilize a ring geometry in order to eliminate strong, confounding effects from symmetry-breaking fields at the edge (Fig.~\ref{fig:Exp_scheme}, A and B). From the power-law decay of spatial correlations, we extract the scaling dimensions, $\Delta_\sigma^{{}^{{}_{\text{1D}}}}$ and $\Delta_\sigma^{{}^{{}_{\text{2D}}}}$,
of the order parameter. Our approach naturally complements recent works that extract scaling dimensions via either holographic digital quantum circuits~\cite{Dborin2022, anand2023holographic, haghshenas2023probing} or the Kibble-Zurek mechanism~\cite{Kibble_1976, Zurek1985, Anquez2016, cui2016, Keesling2019, Ebadi2021, Maxime2022}. Whereas a non-equilibrium approach based on the Kibble-Zurek mechanism is simpler to implement owing to its robustness against decoherence, our adiabatic approach provides full access to all critical exponents. Additionally, the resulting critical state can serve as a quantum resource for further studies. Second, by tuning the amount of decoherence in our Rydberg simulator, we investigate the nature of quantum criticality in open systems. We find that a single, decoherence-induced length scale, $\xi_d$, captures the openness of our quantum system across all parameter regimes~\cite{Maxime2022}. Finally, an additional feature that arises in two dimensions is the possibility that the 1D boundary exhibits its own independent phase transition~\cite{Marcin22}. By varying the interaction strength, we access two distinct boundary universality classes~\cite{AJBray_1977, Burkhardt94, Gliozzi2015}: (i) the ordinary class where the bulk and boundary order simultaneously and (ii) the surface class, where the boundary orders while the bulk remains disordered. 

\section*{Experimental setup}
Our experiment consists of either a 1D ring or a 2D square lattice of $^{133}$Cs atoms trapped in an optical tweezer array (Fig.~1, A and B). The atoms are rearranged from an initial reservoir~\cite{Barredo2016, endres2016} and subsequently Raman sideband cooled~\cite{Liu2019, Kaufman2012} to a low-temperature state with an average motional occupation, $\bar{n} <$ 0.1, for all three axes. We encode a spin-1/2 degree of freedom in the electronic ground state $|g\rangle = |6S_{1/2}, F = 4, m_F = 4\rangle$ and a highly excited Rydberg state $|r\rangle = |54S_{1/2}, J = 1/2, m_J = 1/2\rangle$, where $F$ is the total angular momentum quantum number, $J$ is the total electron angular momentum quantum number, and $m_F, m_J$ are the associated magnetic quantum numbers. A driving field that couples these states is realized via a two-photon transition mediated by the intermediate state, $|7P_{3/2}\rangle$~\cite{Graham19}. The combination of this on-site field and the presence of strong van der Waals interactions between Rydberg atoms, gives rise to the effective Hamiltonian,
\begin{equation} \label{Ham}
H = \frac{\Omega}{2}\sum_i(|g_i\rangle\langle r_i|+|r_i\rangle\langle g_i|) - \Delta \sum_i n_i + \sum _{i<j}V_{ij} n_i n_j,
\end{equation}
where $n_i = |r_i\rangle\langle r_i|$. Here, $V_{ij} = \frac{C_6}{R_{ij}^6}$ characterizes the long-range van der Waals interaction, with $R_{ij}$ being the distance between atoms $i$ and $j$, while $\Omega$ and $\Delta$ correspond to the two-photon Rabi frequency and the detuning of the driving field, respectively.

Parameterizing the Hamiltonian via the ratio of the blockade radius, $R_b = (C_6/\Omega)^{1/6}$, to the lattice spacing $a$, and the ratio, $\Delta / \Omega$, of the detuning to the Rabi frequency (Fig.~\ref{fig:Exp_scheme}C), leads to phase diagrams exhibiting a variety of symmetry-breaking~\cite{Pascal2021,Ebadi2021} and topological orders~\cite{semeghini21}. Our specific focus will be on the continuous phase transition between the disordered paramagnetic phase and the $\mathbb{Z}_2$-ordered antiferromagnetic phase~\cite{Bernien2017, Keesling2019} (Fig.~\ref{fig:Exp_scheme}C), described by the Ising conformal field theory (CFT)~\cite{Fendley04,Slagle2021}. 

\section*{Critical correlations in 1D}

\begin{figure}[t]
   \centering
    \includegraphics[width= 1\linewidth]{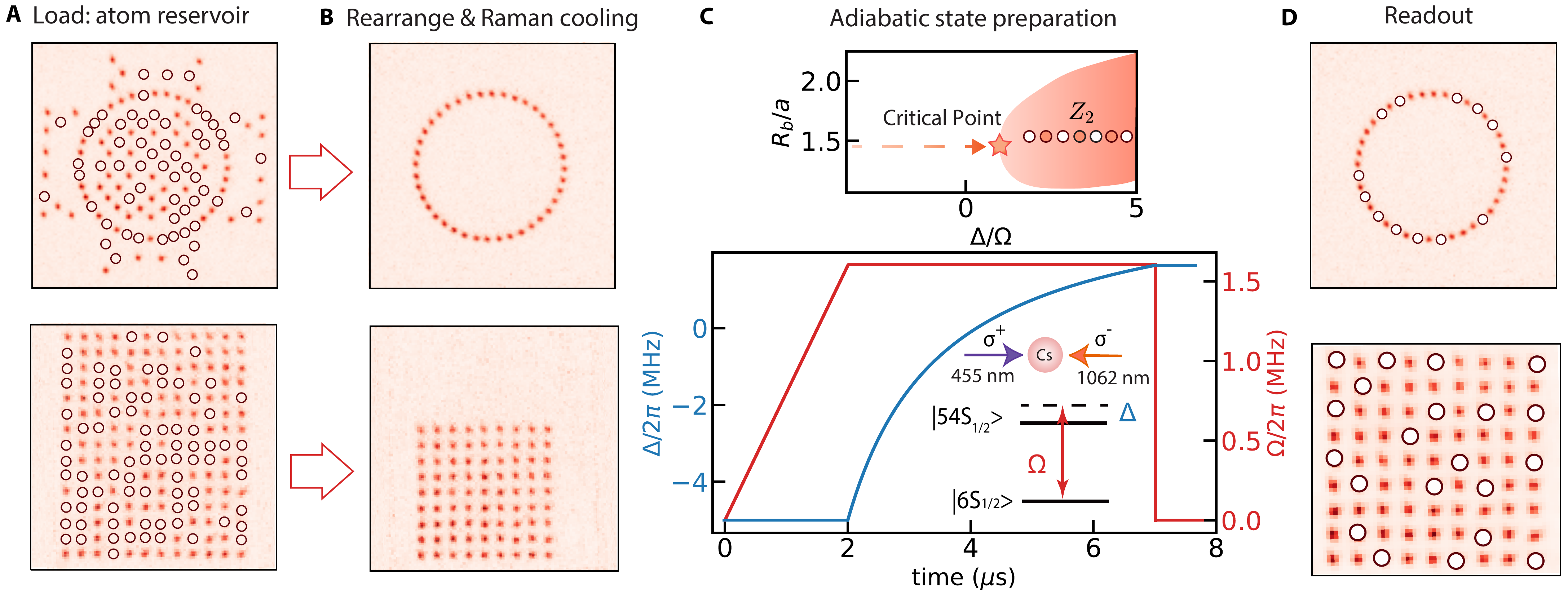}
    \caption{\textbf{Experimental protocols for initialization and state preparation.} (\textbf{A}) Fluorescence images of stochastically-loaded $^{133}$Cs atoms in optical tweezer arrays for 1D (top) and 2D (bottom). Open circles show the locations of unloaded optical tweezers. (\textbf{B}) Defect-free atom arrays after rearrangement. (\textbf{C}) Upper: Schematic of the 1D phase diagram. The dashed arrow denotes the adiabatic state preparation trajectory with the star indicating the critical point. Lower: Optimized ramp profile for adiabatic preparation of the critical state. Here $\Omega$ and $\Delta$ correspond to the two-photon Rabi frequency and the detuning of the driving field, respectively. $R_b$ is the blockade radius. (\textbf{D}) Fluorescence images of the array after adiabatic state preparation. The open circles denote the absence of an atom, which is inferred to be in the Rydberg state.}
    \label{fig:Exp_scheme}
\end{figure}

\begin{figure}
   \centering
    \includegraphics[width= 1\linewidth]{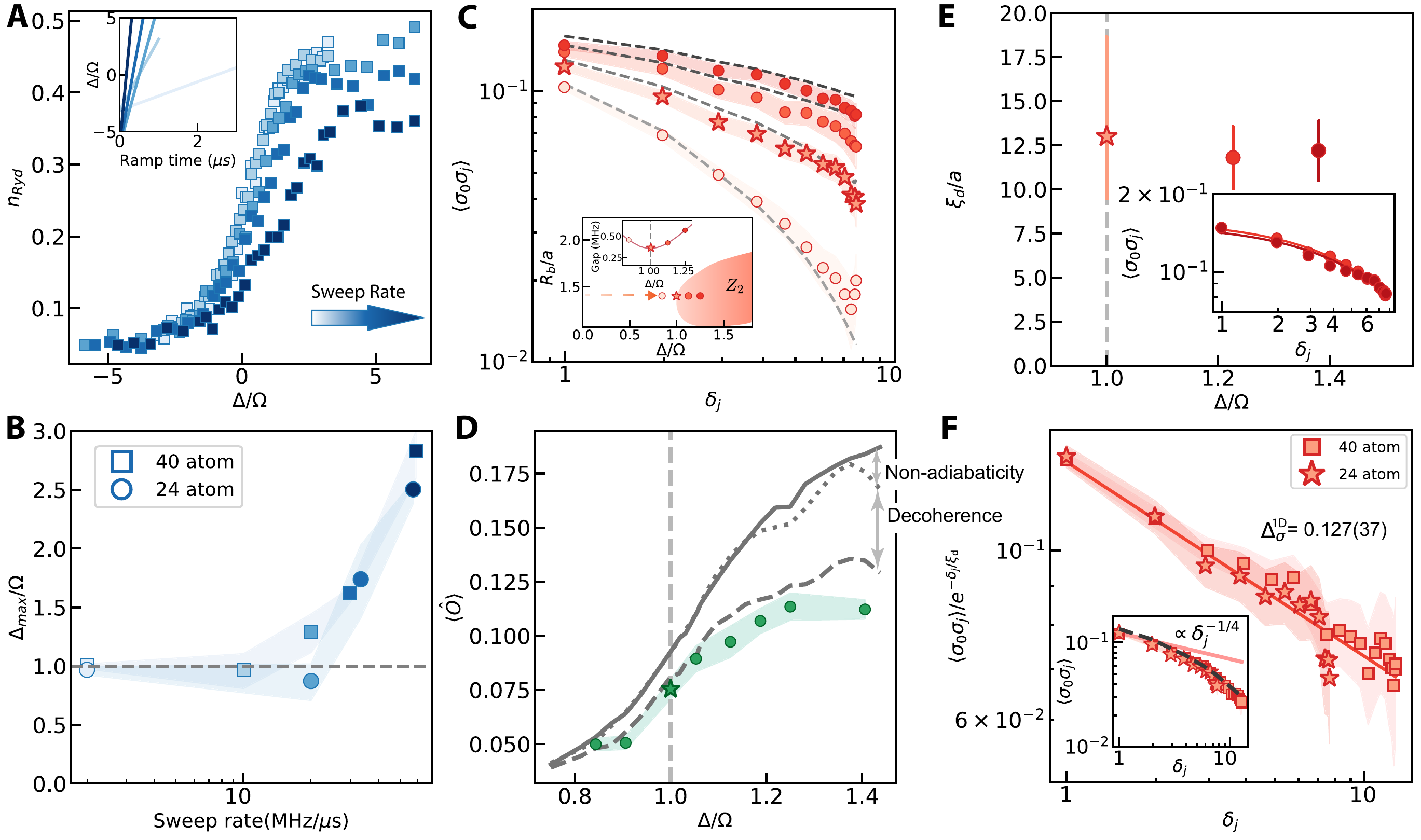}
    \caption{\textbf{Critical correlations in a 1D  ring geometry.} Squares (40-atom), circles (24-atom), stars (24-atom at the critical point) denote experimental measurements, with shaded regions representing 1-$\sigma$ bootstrap errors. Gray solid, dotted, and dashed lines respectively indicate ground state, unitary under optimized ramp, and stochastic wave-function simulation results using experimental parameters. Horizontal and vertical gray lines mark the location of the gap minimum for the 24-atom system. (\textbf{A}) Rydberg population density across the phase transition at different sweep rates. Color gradients indicate the sweep rates variation. Inset: Linear ramp profiles at different sweep rates. (\textbf{B}) Peak location of susceptibility $\chi$ at different sweep rates. (\textbf{C}) $\sigma$ field correlation measurements around the critical point, with red stars representing the critical correlations. Inset: 1D phase diagram and gap profile with red markers denoting the locations of the measurements. (\textbf{D}) Measured order parameter $\langle\hat{O}\rangle$ as a function of the tuning parameter $\Delta/\Omega$. (\textbf{E}) Fitted decoherence-induced length scale $\xi_d/a$ at the critical point and in the AFM phase. The uncertainties in the measurements correspond to a combination of 1-$\sigma$ bootstrap error and fitting error. Inset: Measured $\langle\sigma_0\sigma_j\rangle$ in the AFM phase for two values of $\Delta/\Omega$, with exponential fits to extract $\xi_d/a$. (\textbf{F}) Measured $\langle\sigma_0\sigma_j\rangle$ at critical points divided by $\mathrm{exp}(-\delta_j/\xi_d)$ as a function of $\delta_j$. The results yield a power-law decay with an exponent of 2$\Delta_\sigma^{{}^{{}_{\text{1D}}}}$(solid red line). The uncertainty in $\Delta_\sigma^{{}^{{}_{\text{1D}}}}$ includes both 1-$\sigma$ bootstrap error and fitting error. 
    Inset: Raw measurements of $\langle\sigma_0\sigma_j\rangle$. The solid pink line represents the expected power-law decay for (1+1)d Ising universality class.}
    \label{fig:Critical_Corr}
\end{figure}

To explore the Ising phase transition in 1D, we prepare an $N=24$ atom ring with the lattice spacing chosen such that $R_b/a \approx 1.4$ (Fig.~\ref{fig:Exp_scheme}C). We begin by experimentally locating the critical point, $\Delta_c$, using a Kibble-Zurek-like procedure. At the start of the protocol, all atoms are initialized in the state $|g\rangle$, corresponding to the many-body ground state of the Hamiltonian with $\Omega = 0$ at a large negative detuning. Adiabatically ramping the Rabi frequency then prepares the ground state, after which we linearly sweep $\Delta$ across the phase boundary at different rates. For the slowest sweeps (Fig.~\ref{fig:Critical_Corr}A), one expects the average Rydberg population $\langle n \rangle$ to simply follow that of the instantaneous ground state, with a susceptibility, $\chi = \partial \langle n \rangle / \partial \Delta$, that peaks precisely at the critical point ~\cite{see_sm, Ebadi2021}. For faster sweeps (Fig.~\ref{fig:Critical_Corr}A), the Rydberg population lags behind the ground state expectation and causes the susceptibility to peak after the transition~\cite{Damski20}. As depicted in Fig.~\ref{fig:Critical_Corr}B, we extract the detuning, $\Delta_\mathrm{max}$, corresponding to the peak susceptibility as a function of the sweep rate. We find that $\Delta_{\text{max}}$ converges for sufficiently slow sweeps. We then identify the critical point as $\Delta_c/\Omega$ = 0.97(5) from the $\Delta_\mathrm{max}$ determined at the slowest ramp speed. This value is in agreement with the location of the minimum excitation gap. 

Having identified $\Delta_c$, we now turn to adiabatically preparing the critical ground state. To this end, we further optimize the detuning ramp profile by taking into account the instantaneous energy gap (Fig.~\ref{fig:Exp_scheme}C)~\cite{see_sm}. For our 1D Rydberg array, the primary field $\sigma$ of the emergent Ising CFT is represented, at leading order~\cite{Slagle2021}, by the microscopic lattice operator
\begin{equation}
\sigma_i = (-1)^i(n_i - \langle n\rangle). 
\end{equation}
In the quantum critical ground state, $\langle\sigma_i\rangle = 0$ on each site, but its two-point correlator $\langle \sigma_0\sigma_j \rangle$ decays as a power law,
\begin{equation}
    \langle \sigma_0\sigma_j\rangle \propto \delta_j ^{-2\Delta_\sigma^{{}^{{}_{\text{1D}}}}},
\end{equation}
where $\delta_j/a = \frac{N}{\pi}\sin\left(\frac{\pi j}{N}\right)$ is an effective spatial separation that accounts for the periodic boundary condition~\cite{Slagle2021}. The exponent of this power law is twice the scaling dimension, $\Delta_\sigma^{{}^{{}_{\text{1D}}}}=1/8$, of the (1+1)d Ising CFT.

Interestingly, our data are inconsistent with the power-law decay of correlations in real-space, and instead, appear to fall off significantly more rapidly (Fig.~\ref{fig:Critical_Corr}F, inset). This suggests the presence of a mechanism that inhibits the formation of long-distance correlations. To investigate this, we adiabatically prepare and characterize states away from the critical point, keeping the total preparation time fixed. In the antiferromagnetic (AFM) phase, one analytically expects the correlator, $\langle \sigma_0\sigma_j\rangle$, to exhibit a plateau at large distances, corresponding to long-range order. However, as shown in Fig.~\ref{fig:Critical_Corr}C (dark red circles), we again observe the rapid decay of correlations; in particular, we find that the $\sigma$ field correlator exhibits an exponential decay with length scale $\xi/a$ = 12.0(13) in the AFM phase (inset Fig.~\ref{fig:Critical_Corr}E). This raises the question: what is the microscopic origin of this length scale?

There are two natural possibilities. First, despite our best efforts, the long-distance correlations could still be cut off by diabatic errors. Second, decoherence arising from the openness of our quantum system could also limit the growth of correlations. To distinguish these possibilities, we consider the order parameter 
\begin{equation} \label{OrderParam}
\langle\hat{O}\rangle = \frac{1}{N^2}\sum_{i,j}\langle\sigma_i\sigma_j\rangle,
\end{equation}
where $N$ is the total number of atoms. Compared to the ground state expectation (solid gray curve, Fig.~\ref{fig:Critical_Corr}D), we find that the data (green circles, Fig.~\ref{fig:Critical_Corr}D) exhibit smaller values of the order parameter, with a difference that becomes more pronounced after the critical point. Moreover, we observe that time-dependent simulations (dotted gray line, Fig.~\ref{fig:Critical_Corr}D)~\cite{see_sm}, which account for non-adiabatic errors, yield an order parameter that is quite close to the ground state value. This suggests that the length scale suppressing the correlations originates from decoherence at both the critical point and in the AFM.

Thus, one might naturally expect that the correlations at the critical point are governed by a decay profile of the form~\cite{Maghrebi:2016, Paz:2021b,Maxime2022}, 
\begin{equation} \label{eq:powerlawexp}
    \langle \sigma_0\sigma_j\rangle \propto \delta_j^{-2\Delta_\sigma^{{}^{{}_{\text{1D}}}}} e^{-\delta_j/\xi_\mathrm{d}}.
\end{equation}
This is indeed borne out by the data (inset, Fig.~\ref{fig:Critical_Corr}F). In particular, we extract a decoherence-induced length scale $\xi_\mathrm{d}/a = 13.2_{-3.6}^{+5.7}$, which matches that observed in the AFM (Fig.~\ref{fig:Critical_Corr}E). As depicted in Fig.~\ref{fig:Critical_Corr}F, by accounting for this exponential decay, we observe the characteristic power-law decay of critical correlations, with $\Delta_\sigma^{{}^{{}_{\text{1D}}}} = 0.127(37)$, in agreement with the CFT prediction. 

Three remarks are in order. First, to investigate the robustness of our observed power-law and to ensure a separation of scales between the system size and the decoherence length scale, we prepare and study critical states on a larger $N = 40$ atom ring (Fig.~\ref{fig:Exp_scheme}B). We find that the correlations follow the same functional form, and that the measured scaling dimension and decoherence length scale are identical. Second, although we have focused on decoherence as the dominant source of correlation suppression, one can also access regimes where non-adiabatic effects are manifest ~\cite{see_sm}. Third, we posit that the dominant experimental decoherence mechanisms are \cite{Bluvstein2021, scholl2021quantum, yang2022exploring}: (i) intermediate-state scattering associated with our two-photon excitation scheme and (ii) the finite lifetime of the Rydberg state. We incorporate these effects into a large-scale stochastic wave-function simulation that utilizes independent experimental measurements of these decoherence rates. Notably, this captures---without any additional free parameters---both the decay of correlations (dashed gray, Fig.~\ref{fig:Critical_Corr}C) and the suppression of the order parameter (dashed gray, Fig.~\ref{fig:Critical_Corr}D). 

\begin{figure}
   \centering
    \includegraphics[width= 1\linewidth]{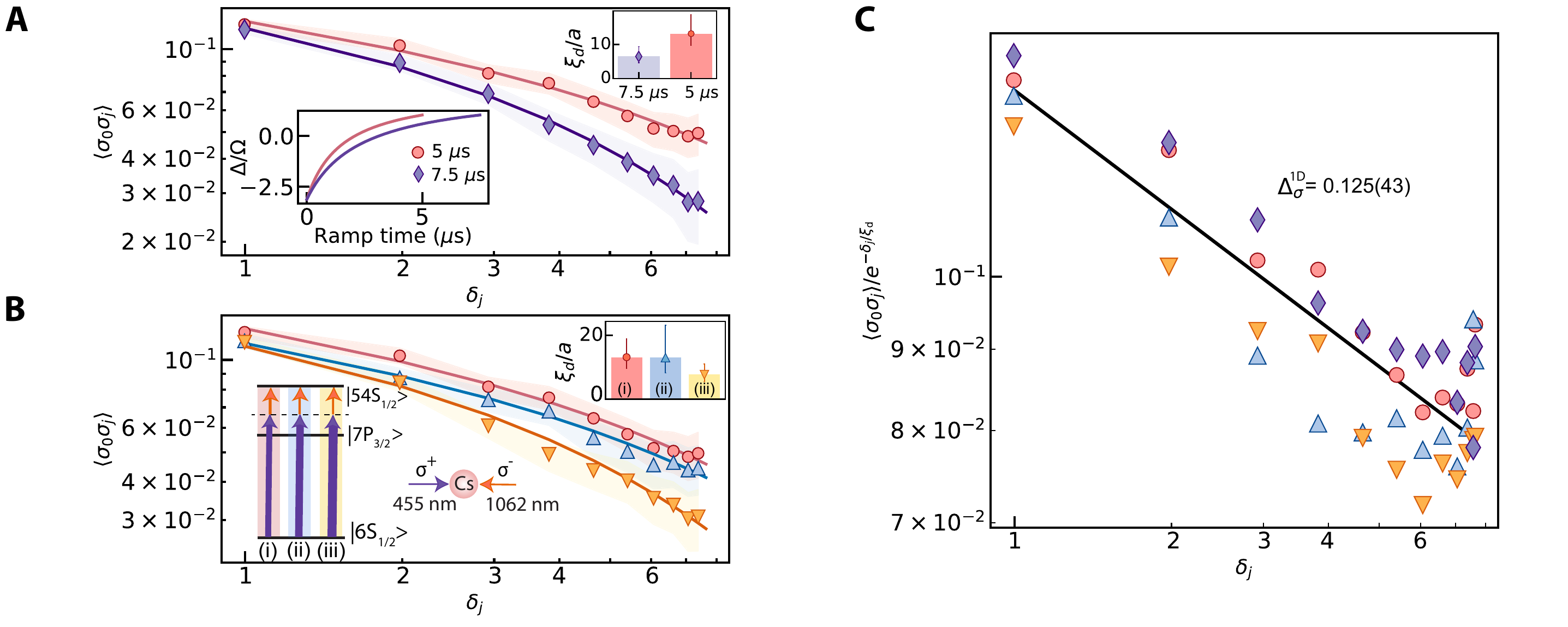}
    \caption{\textbf{Tuning quantum criticality in an open system.} Markers (24-atom) with different shapes are experimental measurements at the critical point under different experimental conditions, with shaded regions denoting the 1-$\sigma$ bootstrap errors. Error bars in fitted $\xi_d/a$ include 1-$\sigma$ bootstrap error and fitting error. (\textbf{A}) $\sigma$ field correlation measurements at two different ramp times. Upper inset: Fitted $\xi_d/a$. Lower inset: Corresponding ramp profiles. (\textbf{B}) $\sigma$ field correlation measurements at three different intensity configurations of Rydberg beams at fixed $\Omega$. Upper inset: Fitted $\xi_d/a$. Lower inset: A schematic showing the single photon rabi frequencies $\Omega_{455}$ and $\Omega_{1062}$ for each configuration. (\textbf{C}) Correlation measurements in (A) and (B) divided by the exponential $\mathrm{exp}(-\delta_j/\xi_d)$, displaying a collapse onto a power-law decay with an exponent of 2$\Delta_\sigma^{{}^{{}_{\text{1D}}}}$ (solid black line), acquired via a simultaneous fit to the power-law exponential model for all scenarios. The uncertainty in $\Delta_\sigma^{{}^{{}_{\text{1D}}}}$ includes both 1-$\sigma$ bootstrap error and fitting error.}
    \label{fig:SigCorr_24vs40}
\end{figure}

\textbf{Tuning open system criticality}---Our observations suggest that the principal effect of decoherence on quantum criticality is to introduce a single length scale into an otherwise scale-invariant state~\cite{Maxime2022}. This provides a simple conceptual framework for extracting the universal scaling dimensions from open quantum systems at criticality. To investigate the generality of this framework, we directly tune the amount of decoherence in our system, by either increasing the total ramp time (Fig.~\ref{fig:SigCorr_24vs40}A) or the intermediate-state scattering rate (Fig.~\ref{fig:SigCorr_24vs40}B). In all cases, the correlations decay as in Eq.~\ref{eq:powerlawexp}, but with different decoherence length scales (Fig.~\ref{fig:SigCorr_24vs40}, A and B). Despite a relatively large range of values for $\xi_d$, the extracted scaling dimension remains unchanged. This is evinced by the collapse of the data onto the universal Ising CFT power-law (across both ramp times and scattering rates) once $\xi_d$ is accounted for (Fig.~\ref{fig:SigCorr_24vs40}C). 

\section*{Critical Correlations in 2D}
\begin{figure}
   \centering
    \includegraphics[width= 1\linewidth]{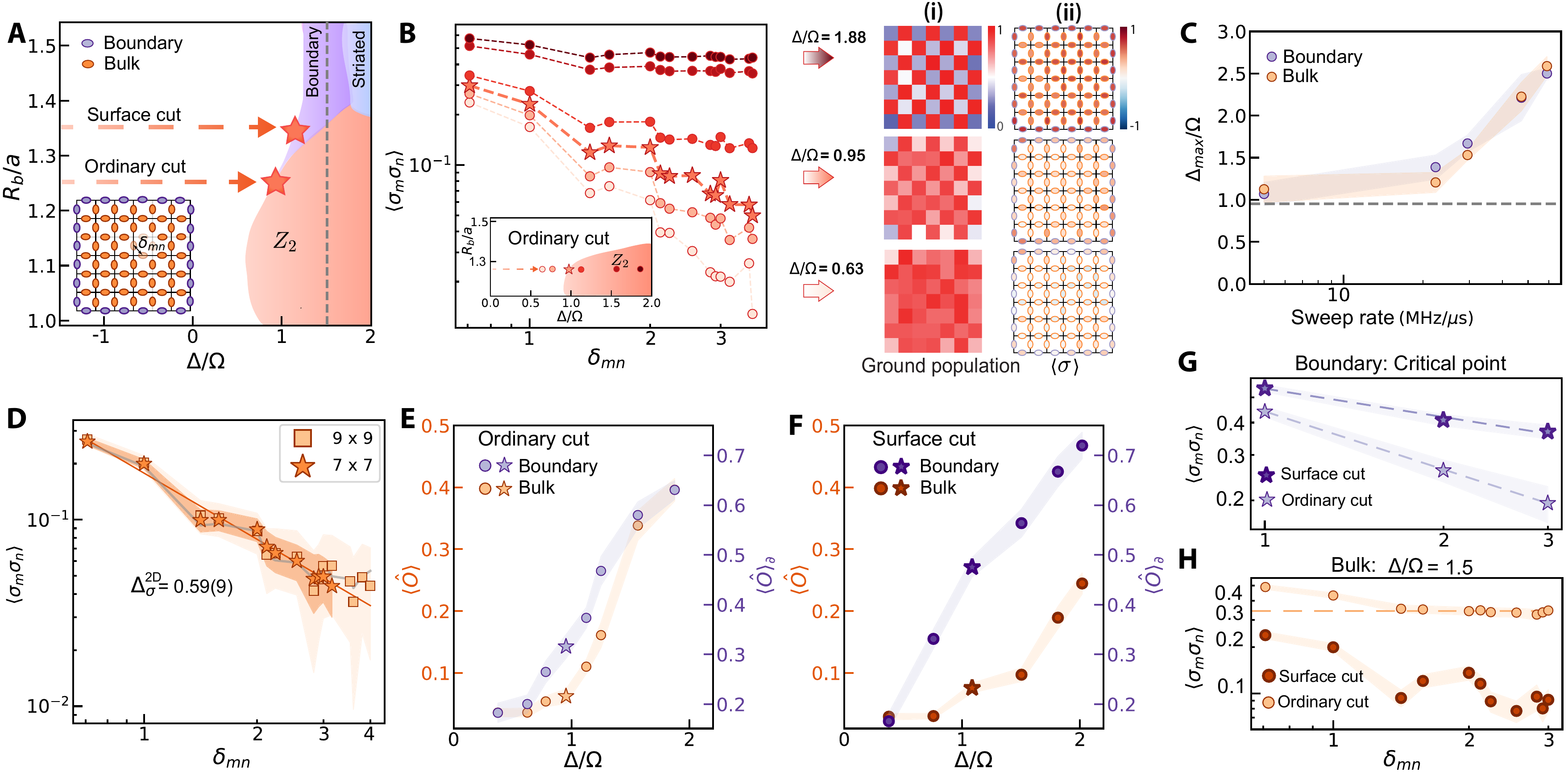}
    \caption{\textbf{Probing critical Ising correlations and surface transitions in a 2D square lattice.} Circles (7 x 7), stars (7 x 7 at the critical point) and squares (9 x 9) denote experimental measurements, with shaded regions representing 1-$\sigma$ bootstrap errors. (\textbf{A}) A schematic for the 2D phase diagram, adapted to the finite system size. The shaded area labeled boundary marks the region of a boundary-ordered phase with a disordered bulk. Red stars indicate the location of the critical point along the two cuts. Gray dashed line indicates $\Delta/\Omega = 1.5$. Inset: The $\sigma$ fields live on lattice bonds, both along the boundary (purple) and within the bulk (orange). $\delta_{mn}$ denotes the Euclidean distance between two nearest $\sigma$ fields. (\textbf{B}) $\sigma$ field correlation measurements around the critical point along the ordinary cut, with measured ground state population (i) and $\sigma$ field (ii) across the phase transition. The critical spatial correlation is highlighted by the star marker. Dashed lines are guides to the eye. Inset: 2D phase diagram schematic with markers indicating the measurements' locations. (\textbf{C}) Peak location of susceptibility $\chi$ along the ordinary cut at different sweep rates with boundary and bulk analyzed separately. Gray dashed line marks the location of gap minimum. (\textbf{D}) The measured critical $\sigma$ field correlation within the bulk exhibits a power-law decay with an exponent of 2$\Delta_{\sigma}$, shown by the orange solid line. The gray solid line represents the ground state simulation using experimental parameters. (\textbf{E}, \textbf{F}) Order parameter analyzed independently for the bulk ($\langle \hat{O} \rangle$) and the boundary ($\langle\hat{O}\rangle_\partial$) across the phase transition along the ordinary cut (E) and the surface cut (F). (\textbf{G}, \textbf{H}) Measured $\sigma$ field spatial correlations along the boundary (G) at the critical point for the two cuts (marked by red stars in (A)), and within the bulk (H) at $\Delta/\Omega$ = 1.5 (marked by the gray dashed line in (A)). Dashed lines are guides to the eye.}
    \label{fig:SigCorr_2D}
\end{figure}

We now turn to the exploration of quantum criticality in two spatial dimensions. Working with a square lattice and a blockade radius of $R_b/a=1.25$ (Fig.~\ref{fig:SigCorr_2D}A), our model exhibits two phases analogous to the 1D case: at low detuning, a trivial paramagnet, and at high detuning, a $\mathbb{Z}_2$ checkerboard state that spontaneously breaks translation symmetry~\cite{Fendley04, Bernien2017}. The transition between these two phases is in the (2+1)d Ising universality class. Its primary $\sigma$ field can be associated with a microscopic operator residing on the lattice bonds (inset, Fig.~\ref{fig:SigCorr_2D}A). Namely, between two neighboring lattice sites at coordinates $(x_i, y_i)$ and $(x_j, y_j)$, the operator is given by
\begin{equation}
  \sigma_{i,j} = (-1)^{x_i+y_i}(n_{i} - n_{j}). \label{eq:sigma2D}
\end{equation} 
We begin by utilizing optimized ramps to prepare the ground state (on a 7$\times$7 array) at various detunings (inset, Fig.~\ref{fig:SigCorr_2D}B). Owing to the odd-length, open-boundary condition geometry, there is a unique ground state in the ordered phase with $\langle \sigma \rangle \approx +1$ [panel (ii) in Fig~\ref{fig:SigCorr_2D}B], which exhibits a checkerboard pattern of $\langle n_i\rangle$ [panel (i) in Fig~\ref{fig:SigCorr_2D}B].

Next, we employ the same Kibble-Zurek-like procedure to experimentally locate the critical point, finding $\Delta_c/\Omega = 1.03(11)$ (Fig.~\ref{fig:SigCorr_2D}C). The (2+1)d Ising CFT predicts that correlations at the critical point should decay as $\langle \sigma_m\sigma_n\rangle \propto \delta_{mn} ^{-2\Delta_{\sigma}^{2D}}$, where $\delta_{mn}$ is the Euclidean distance between bond centers $m$ and $n$. Unlike in the 1D case, the 2D critical exponent $\Delta_\sigma^{2D}\approx$ 0.518149 is not known exactly, but has been estimated to high precision using  conformal bootstrap techniques~\cite{Kos2016}.

We adiabatically prepare the critical ground state for both $7\times 7$ and $9\times9$ square arrays. In this setting, we are able to directly observe real-space power-law correlations unimpeded by decoherence (Fig.~\ref{fig:SigCorr_2D}D). This is enabled by two factors: (i) the larger scaling dimension causes power-laws to manifest at shorter distances, and (ii) 
the previously-extracted decoherence length scale, $\xi_\mathrm{d} \approx 13$, exceeds our linear system size. A simultaneous fit of both system sizes yields $\Delta_\sigma^{{}^{{}_{\text{2D}}}}= 0.59(9)$; in order to minimize boundary effects, we only include $\sigma$ fields within the bulk (inset, Fig.~\ref{fig:SigCorr_2D}A). Although our measured scaling dimension is slightly larger than the CFT prediction, 
it agrees with ground state DMRG calculations. In fact, the correlation function matches the simulations within statistical error for all distances (Fig.~\ref{fig:SigCorr_2D}D); this suggests that the discrepancy with the CFT prediction is a finite-size effect.

\textbf{Boundary phase transitions}---Thus far we have focused on bulk criticality, but the boundary itself can also exhibit rich physics in two dimensions. At $R_b/a=1.25$, we observe that the bulk and boundary order simultaneously (Fig.~\ref{fig:SigCorr_2D}C)---a so-called
ordinary transition~\cite{AJBray_1977, Burkhardt94, Gliozzi2015}. However, this is not the only possibility. In an alternative case, known as a surface transition, the boundary orders independently prior to the bulk phase transition~\cite{AJBray_1977}; for example, this is expected to occur when $R_b/a = 1.35$~(Fig.~\ref{fig:SigCorr_2D}A)~\cite{Marcin22}. 

To investigate this, we prepare states at various detunings along both the ordinary ($R_b/a = 1.25$) and surface ($R_b/a = 1.35$) cuts depicted in Fig.~\ref{fig:SigCorr_2D}A. We measure the order parameter  (Eq.~\ref{OrderParam}), but now separately average over bonds either within the bulk, $\langle \hat{O} \rangle$, or at the boundary, $\langle \hat{O} \rangle_\partial$. Along the surface cut (Fig.~\ref{fig:SigCorr_2D}F), the growth of  $\langle\hat{O}\rangle$ exhibits a discernible lag behind $\langle\hat{O}\rangle_\partial$, particularly when compared to the growth observed along the ordinary cut (Fig.~\ref{fig:SigCorr_2D}E). This is consistent with the expected surface transition~\cite{Marcin22}.

A direct comparison of the spatial correlations along the  cuts further illustrates the influence of the two transitions (Fig.~\ref{fig:SigCorr_2D}, G and H). First, we examine the boundary correlations, $\langle\sigma_m\sigma_n\rangle_\partial$, at the ordinary and surface critical points (stars, Fig.~\ref{fig:SigCorr_2D}A). At the surface transition, $\langle\sigma_m\sigma_n\rangle_\partial$ exhibits a slower decay with larger absolute values compared to the ordinary transition; this is in qualitative agreement with the theoretical prediction that the scaling dimension of the surface transition is significantly smaller than that of the ordinary transition~\cite{see_sm}. However, our ability to quantitatively determine this scaling dimension is constrained by the presence of strong corner effects. 

For comparison, we also analyze the bulk correlations $\langle\sigma_m\sigma_n\rangle$ along $\Delta/\Omega$ = 1.5 (dashed gray, Fig.~\ref{fig:SigCorr_2D}A), which lies within the ordered phase for the ordinary cut, but should only have boundary order for the surface cut. Indeed, as shown in Fig.~\ref{fig:SigCorr_2D}H, the ordinary cut exhibits a plateau in the spatial correlations at large distances indicative of an ordered bulk, whereas the surface cut exhibits rapidly decaying correlations.

\section*{Discussion and Outlook}
Looking forward, our work lays the foundation for several directions. First, producing larger square arrays and improving coherence times should allow direct access to various boundary universality classes (surface, ordinary, and extraordinary) in experiment. Second, bulk ground state physics is enriched by the presence of geometric frustration: odd-length rings support a $W$-state in the ordered phase~\cite{Jovan23, Ruben23}, and in two dimensions one can find 3D XY critical points---with an associated extraordinary-log boundary universality class~\cite{Padayasi:2022}---as well as transitions into gapless and topological spin liquids~\cite{semeghini21}; each of of these should be possible to study using the same adiabatic preparation framework. Finally, quantum critical states are expected to display rich dynamics when out of equilibrium~\cite{Calabrese:2016,Berdanier:2017}, including holographic signatures of emergent gravitational physics \cite{sahay2024emergent}. Such topics have been extensively studied theoretically in (1+1)d CFTs---the extension to (2+1)d is non-trivial, and would benefit greatly from experimental study.

\bibliographystyle{sciencemag}
\bibliography{main}
\section*{Acknowledgements}
We gratefully acknowledge the insights of and discussions with J.E.~Moore, M.~Endres, C.R.~Laumann, Y.~Bao, B.~Ye, P.~Osterholz, R.~Verresen, D.~Mark and S.~Choi. We thank Y.~Yu for contributions to the development of the  control system that enables atom rearrangement. 
\paragraph*{Funding:}
This material is based upon work supported by the U.S. Department of Energy, Office of Science, National Quantum Information Science Research Centers, Quantum Systems Accelerator (award number 7568717). K.W. is supported by an NSF GRFP fellowship. J.W. acknowledges support by the Department of Energy Computational Science Graduate Fellowship under award number DE-SC0022158.
\paragraph*{Author contributions:}
$^\dagger$F.F., K.W., V.S.L., and Y.W. contributed equally to this work. F.F., K.W., Y.W., and R.C. contributed to building the experimental set-up, performed the measurements, and analyzed the data. V.S.L, J.W., M.B., and A.P. conducted the theoretical analysis and simulations. J.K., K.K.N., and N.Y.Y. supervised the work. All authors discussed the results and contributed to the manuscript. 
\paragraph*{Competing interests:}
The authors declare that they have no competing interests.
\paragraph*{Data and materials availability:}
All data needed to evaluate the conclusions in the paper are present in the paper or the supplementary materials. All data presented in this paper are available at Zenodo~\cite{zenodo}.
\subsection*{Supplementary materials}
Materials and Methods\\
Supplementary Text\\
Figs. S1 to S10\\
References \textit{(54-68)}
\clearpage

\clearpage
\renewcommand{\thefigure}{S\arabic{figure}}
\renewcommand{\thetable}{S\arabic{table}}
\renewcommand{\theequation}{S\arabic{equation}}
\renewcommand{\thepage}{S\arabic{page}}
\setcounter{figure}{0}
\setcounter{table}{0}
\setcounter{equation}{0}
\setcounter{page}{1} 
\def\scititle{Probing critical phenomena in open quantum systems using atom arrays}
\begin{center}
\section*{Supplementary Materials for\\ \scititle}
Fang Fang$^{1,2,3\dagger}$, Kenneth Wang$^{1,2,3\dagger}$, Vincent~S.~Liu$^{2\dagger}$, Yu Wang$^{1,2,3\dagger}$,\\ Ryan Cimmino$^{1,2,3}$, Julia Wei$^{2}$, Marcus Bintz$^{2}$, Avery Parr$^{2}$,\and Jack Kemp$^{2}$, Kang-Kuen Ni$^{1,2,3\ast}$, Norman~Y.~Yao$^{2,1,3\ast}$\\
\small{$^\ast$Corresponding author. E-mail: ni@chemistry.harvard.edu, nyao@fas.harvard.edu}
\\
\small{$^\dagger$These authors contribute equally}
\end{center}
\subsubsection*{This PDF file includes:}
Materials and Methods\\
Supplementary Text\\
Figs. S1 to S10\\
References 
\newpage
\tableofcontents
\newpage
\subsection*{Experimental system \label{Sec:expt}}
\addcontentsline{toc}{section}{Experimental system}
\subsubsection*{Optical tweezer arrays}
\addcontentsline{toc}{subsection}{Optical tweezer arrays}
We load optical tweezer arrays of up to 180 Cs atoms from a magneto-optical trap with a loading probability of $40-60\%$ per site. The tweezer traps have a loading trap depth of about $2\pi \times 13$ MHz with radial and axial trap frequencies of about $2\pi \times 90$ kHz and $2\pi \times 14$ kHz, respectively.

We generate arbitrary-geometry tweezer arrays with a Hamamatsu X15213-03R spatial light modulator (SLM) focused through a Jenoptik microscope objective (NA = 0.55). The SLM applies a phase hologram to the laser wavefront, calculated using the phase-fixed weighted Gerchberg-Saxton (WGS) algorithm \cite{Kim19}. We correct for optical aberrations by adding Zernike polynomials with variable amplitudes to the wavefront \cite{Ebadi2021}. This process increases the trap depth by approximately 10\%. We further improve trap depth uniformity to under 2\% (standard deviation/mean) through the adaptive-WGS algorithm. The SLM is illuminated by up to 5 W of 1064 nm light generated by a fiber amplifier (Precilasers, YFA-SF-1050-50-CW) seeded by a low-noise narrow-line Coherent Mephisto MOPA laser. 

After stochastically loading the SLM tweezer array, we use mobile tweezers at 1038 nm created by crossed AA Optoelectronics DTSX-400-1030 acousto-optical deflectors (AODs) to rearrange atoms into a defect-free pattern of our choice. The rearrangement tweezers are generated by a fiber amplifer (Azur Light Systems, ALS-IR-1040-20-A-SF) seeded by a home-build external cavity diode laser using an Innolume gain chip (GM-1030-130-PM-200).

\subsubsection*{Rearrangement}
\addcontentsline{toc}{subsection}{Rearrangement}
Atoms are imaged via fluorescence imaging on our Andor electron-multiplying charge-coupled-device (EMCCD) camera, after which the images are processed and the correct moves are determined. During rearrangement, we grab atoms using AOD tweezers at approximately twice the trap depth of the static SLM tweezers. The AOD tweezers are turned on and off in about $ 200 \text{ } \mu\text{s}$ and moved at about $ 110 \text{ }\mu\text{m}/\text{ms}$. 

We design our rearrangement algorithms to operate on grids, a natural capability of AODs. The sequence of atom moves and ejections must be tailored to the geometry of the SLM tweezers. Each SLM array includes both `target' sites belonging to the desired defect-free pattern and `reservoir' sites supplying additional atoms to fill unloaded target sites. In both 1D and 2D, we are careful to minimize the number of grab and drop events of the atoms.

\begin{figure}
   \centering
    \includegraphics[width= 1\linewidth]{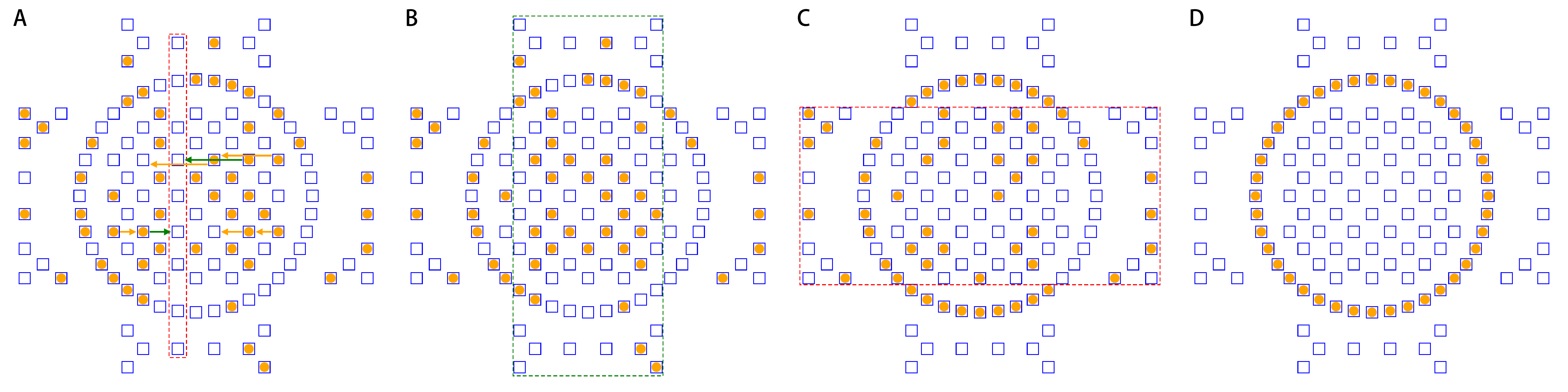}
    \caption{\textbf{Rearrangement procedure in 1D for a 40-atom ring}. Blue squares denote locations of SLM traps, and orange circles denote the locations of atoms. Each target site on the ring is associated with a reservoir row or column. 
    (\textbf{A}) Row by row, we shuffle atoms horizontally to columns that need additional atoms to reach their targets. Green arrows denote atoms that are moved in order to fill the deficient red column. Orange arrows denote atoms that are moved to more central columns to contribute more atoms to the next step. We note that if all columns have a sufficient number of atoms, this step can be skipped. Furthermore, not all rows will be moved. (\textbf{B}) Column by column (within the green rectangle), we shuffle atoms vertically to their target sites, ejecting any unneeded atoms and also filling rows that need additional atoms. Importantly all atoms are now in the central rows which will be targeted in the next step. (\textbf{C}) Row by row (within the red rectangle), we place the remaining atoms into the target sites, while ejecting unneeded atoms. (\textbf{D}) After the previous rearrangement steps, a defect free ring is achieved.}
    \label{fig:rearr-circle}
\end{figure}

\textit{1D rearrangement procedure.} For our 1D experiments on a ring, we load additional atoms into rows and columns of traps situated around and inside the ring to serve as a reservoir. This reservoir is designed to maximize connectivity along a row/column, reduce redundancies, and enforce a minimum spacing between traps (see Fig.~\ref{fig:rearr-circle}). While the AOD is capable of simultaneously producing multiple tweezers, we only use a single mobile tweezer to rearrange atoms on a ring. This minimizes atom heating due to beat-note frequencies of adjacent tweezers moving over a non-uniform grid, which would otherwise cause trap depth modulations on the order of our trap frequencies.

\begin{figure}
   \centering
    \includegraphics[width= 1\linewidth]{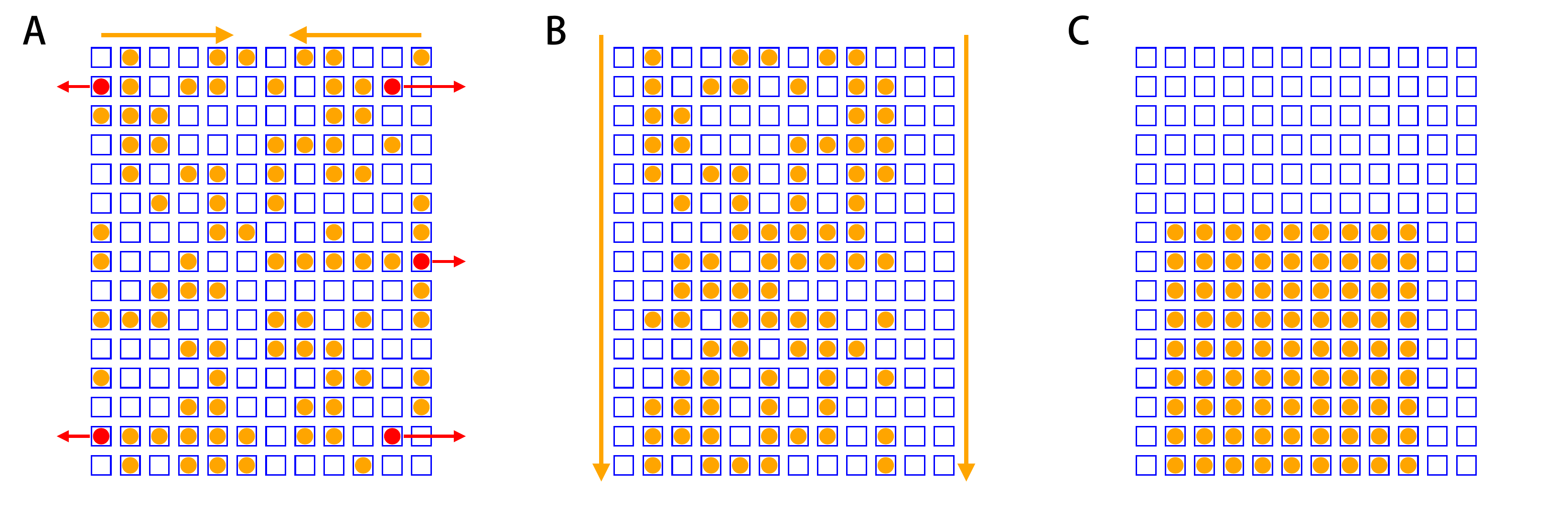}
    \caption{\textbf{Rearrangement procedure in 2D for a 9x9 rectangular array}. Blue squares denote locations of SLM traps, and orange circles denote the locations of atoms.
    (\textbf{A}) Row by row, we shuffle atoms horizontally to columns that need additional atoms and eject unnecessary atoms (red arrows). Note that atoms in the final target region are moved as little as possible. (\textbf{B}) We move all atoms downward to the bottom of the array. (\textbf{C}) After the previous rearrangement steps, a defect free rectangular array is achieved.}
    \label{fig:rearr-rect}
\end{figure}

\textit{2D rearrangement procedure.} For our 2D experiments, we rearrange over a rectangular array. Because of the regular geometry, atoms of a given row or column can be simultaneously rearranged by AOD tweezers without low frequency beatnotes, allowing us to move atoms in parallel without substantial heating (see Fig.~\ref{fig:rearr-rect}).

\subsubsection*{Rydberg laser system}
\addcontentsline{toc}{subsection}{Rydberg laser system}
We coherently excite atoms to the Rydberg state using a two-photon transition via the intermediate state $|7P_{3/2}\rangle$, with counter-propagating laser beams at $455 \text{ nm}$ and $1062 \text{ nm}$. The Rydberg excitation beams are shaped into elliptical Gaussian beams to improve their intensity uniformity across the array, with $(\omega_x, \omega_y)_{455}$ = (87, 137) $\mu$m and $(\omega_x, \omega_y)_{1062}$ = (64, 208) $\mu$m at the position of the atoms. In a typical experiment, the powers of each laser are chosen to achieve single-photon Rabi frequencies of $(\Omega_{455}, \Omega_{1062}) \approx 2\pi \times (80, 42)$ MHz. We operate at an intermediate state detuning of 1.06 GHz, and the two-photon Rabi frequency ranges from 1 to 1.6 MHz.

The 455 nm laser is a frequency-doubled Ti:Sapphire laser from M Squared. The fundamental frequency is locked to an ultra-low-expansion (ULE) cavity (Notched cavity from Stable Laser Systems) with finesse $\mathcal{F}$ = 24000 at 911 nm. The 1062 nm laser is an external cavity quantum dot laser (Time-base ECQDL-200FC) locked to the same ULE cavity with finesse $\mathcal{F}$ = 26000 at 1062 nm. To minimize the phase noise of the 1062 nm laser, we inject the cavity-filtered transmitted light of the ULE cavity (57 kHz linewidth), into a laser diode (EYP-RWL-1060-00100-1300-SOT01-0000). The output of this laser diode is further amplified by a 50 W fiber amplifier (Precilasers, YFA-SF-1050-50-CW). Phase noise suppression enhances the single-atom ground-Rydberg Rabi coherence time from 4 $\mu$s to approximately 20 $\mu$s. 

We use acousto-optical modulators (AOM) to dynamically change the detuning and power of each laser. We change the 455 nm laser power and the two-photon detuning using an AOM (AA Optoelectronics, MQ180-A0.25-VIS, 43 MHz maximum modulation bandwidth) in double-pass configuration, while the power of the 1062 nm laser is changed with a single-pass AOM from Gooch and Housego. During a Rydberg excitation sequence, the 1062 nm laser is held at constant amplitude and frequency while the 455 nm laser amplitude and frequency are varied. To achieve the stringent bandwidth and response time requirements of our experiment, we focus the 455 nm
beam to 100 $\mu$m diameter at the AOM aperture, resulting in an optical response time of 16 ns. Our fastest ramp lasted about 350 ns, during which we swept the detuning by 20 MHz—equivalent to a 10 MHz shift on the double-passed AOM. It is well within the limitations exerted by
both the RF bandwidth (43 MHz) and optically-limited response time (16 ns). Both lasers are intensity-stabilized using feedback from a photodiode and a home-built proportional-integral-derivative circuit. Due to the short pulse length for the blue laser, a sample and hold circuit is additionally used for intensity stabilization, with a sample period that occurs within 20 ms of the pulse. Nevertheless, we observe a long-term drift in the two-photon Rydberg Rabi frequency, possibly caused by slow drifts in laser polarization and pointing. This results in a systematic error, bounded by 5\%, in the critical point's location.

\subsection*{Adiabatic state preparation}
\addcontentsline{toc}{section}{Adiabatic state preparation}
\subsubsection*{State initialization}
\addcontentsline{toc}{subsection}{State initialization}
Following rearrangement, we optically pump atoms to the ground state $|6S_{1/2}, F=4, m_F=4\rangle$ in an 8.8 G magnetic field. We perform 3D Raman sideband cooling to reduce the average motional quantum number $\overline{n}$ to less than 0.1~\cite{Liu2019}, taking approximately 100 ms. Following Raman cooling, we adiabatically lower the trap in about 100 $\mu$s before turning it off completely during the Rydberg excitation. Although releasing the atom from a lower trap depth results in lower kinetic energy, the position distribution is wider resulting in greater sensitivity to static positional disorder. However, at a higher trap depth, releasing the atom results in greater kinetic energy, so the atom will move more during the Rydberg excitation. The optimal trap depth for release is numerically determined to be $2\pi \times 1.4$ MHz with radial and axial trap frequencies of $2\pi \times 30$ kHz and $2\pi \times 4.7$ kHz, respectively, by minimizing the atom position distribution $\sigma_r + \sigma_v \times t_{evolve}$ during state evolution, where $\sigma_r$ and $\sigma_v$ is the spread of the initial atom position and momentum upon trap release.

\subsubsection*{State evolution - Local linear adiabatic (LILA)
ramp}
\addcontentsline{toc}{subsection}{State evolution - Local linear adiabatic (LILA)
ramp}
Consider a system with the Rydberg Hamiltonian (Eqn. \ref{Ham} in the main text) that is evolving in time according to a time-dependent detuning $ \Delta(t) $ that starts at $\Delta(0) = \Delta_0$. According to the adiabatic theorem, the system will remain in the ground state of $H(t)$ at a later time $t$, provided that the evolution of the Hamiltonian is slow enough to satisfy \cite{richerme2013experimental, Amin09, Roland02}
\begin{equation}
   \min_{t \in [0, T]} \Big|\frac{E_g(t)^2}{\dot{\Delta}(t)} \Big| \gg 1.
\end{equation}
Here $E_g(t)$ denotes the time-dependent gap between the ground state and the first excited state of the instantaneous $H(t)$. Our goal is to maximize $\min\limits_{t \in [0, T]} \gamma(t)$, where $\gamma(t) = \frac{E_g^2}{\dot{\Delta}}$ for a fixed time $ T $, which is realized when $\gamma(t) = \gamma$ is a constant. Therefore, our task is to find a ramp profile $\Delta(t)$ that fulfills the condition
\begin{equation}
 \frac{d\Delta}{dt} = \frac{E_g^2(\Delta(t))}{\gamma}.
\end{equation}
with two boundary conditions $\Delta(0) = \Delta_0$ and $\Delta(T) = \Delta_c$. We solve this problem on a equal-spacing discretized grid in detuning with $ N $ points. We set $\Delta_{k=0} = \Delta_0$ and $\Delta_{k=N} = \Delta_c$, and $(\Delta_{k+1} - \Delta_{k})T \ll 1$, where $ T $ is the total time. The corresponding time spacings can be solved for, and we obtain 
\begin{align}
\Delta t_k &= \frac{T}{E_g^2(\Delta_k)\sum\limits_{k=0}^{n} 1/E_g^2(\Delta_k)}, \\
\gamma &= \frac{T}{d\Delta\sum\limits_{k=0}^{n}1/E_g^2(\Delta_k)}.
\end{align}
This formula allows us to numerically evaluate an optimal ramp profile $\Delta(t)$ given a known gap profile $E_g(\Delta)$, an initial detuning $\Delta_0$ and final detuning $\Delta_c$, and a fixed total time $T$. 

However, this solution requires knowledge of the gap at all detunings, which can be computationally expensive. To reduce the computational cost, we approximate the gap profile as linear around the critical detuning $\Delta_c$: 
\begin{equation}
E_g(\Delta) = E_0 + \frac{E_c-E_0}{\Delta_c-\Delta_0}(\Delta-\Delta_0).
\end{equation} 
This results in an analytic solution:
\begin{align}
  \Delta(t) &= \frac{E_0\Delta_c t+E_c\Delta_0(T-t)}{E_0t+E_c(T-t)}, \\
  \gamma &= \frac{E_0E_cT}{\Delta_c-\Delta_0}.
\end{align}
which is the optimized ramp profile that is used throughout our experiments.  

\subsubsection*{Rydberg detection}
\addcontentsline{toc}{subsection}{Rydberg detection}
To detect the state of our atoms, we apply a microwave pulse \cite{Graham19, Ebadi2021} at approximately $ 9.16 \text{ GHz}$, followed by a rapid turn on of the tweezer traps, expelling atoms in the Rydberg state. Ground state atoms are recaptured, so subsequent imaging of these atoms provides a measurement of the Rydberg population of each site. There are two sources of error in this procedure. First, a false positive error, where a ground state atom has been lost, and is misinterpreted as a Rydberg atom. Second, a false negative error, where a Rydberg atom has decayed to the ground state and is interpreted as a ground state atom. We address the first error by not analyzing snapshots with blockade violations (See later Section on data post-selection). For the second error, we either apply a single Rydberg $\pi$ pulse or two Rydberg $\pi$ pulses before state detection, to separate the contribution from the false negative detection error and the $\pi$ pulse error. 

The detailed procedure is outlined as follows. We first define three quantities:
\begin{itemize}
 \item 1-$\eta_0$: False positive detection error where a ground state atom is detected as a Rydberg atom.
 \item p: Rydberg excitation probability after a $\pi$ pulse.
 \item $\epsilon$: False negative detection error where a Rydberg atom is detected as a ground state atom.
\end{itemize} 
We determine $\eta_0$ = 0.980(8) via a release-recapture experiment. To further determine $\epsilon$, we additionally perform two experiments. In the first experiment, we apply a Rydberg $\pi$ pulse and then detect the ground state population $n_{g1}$ = 0.053(14). In the second experiment, we apply two Rydberg $\pi$ pulses and then do the same detection for the ground state population $n_{g2}$ = 0.86(1). With the three quantities defined above, we end up with two equations:
\begin{equation}
    \eta_0(1-p) + \epsilon p = n_{g1}
\end{equation}

\begin{equation}
    \eta_0(p^2 + (1-p)^2) + 2\epsilon p(1-p) = n_{g2}
\end{equation}
We then solve the above equations to estimate $\epsilon$. The errors in $\epsilon$ are done by propagating the standard deviation of the known quantities $\eta_0$, $n_{g1}$ and $n_{g2}$. This calibration procedure yields our estimate of the error, which is less than 0.015. We note that to achieve a good calibration of $\eta_0$, $n_{g1}$ and $n_{g2}$, the above experiments are repeated by more than 10000 times. The simulation results presented in this work do not include the false negative error correction. 

\subsection*{Additional Sources of Imperfection}
\addcontentsline{toc}{section}{Additional Sources of Imperfection}
\subsubsection*{Non-adiabaticity}
\addcontentsline{toc}{subsection}{Non-adiabaticity}
When the ramp speed exceeds the time scale determined by the excitation gap, a finite density of excitations above the many-body ground state becomes populated. Energy gaps decrease in system size as $1/L$ for 1D \cite{Henkel_1987}, necessitating a longer time scale to ensure adiabaticity in larger systems. The challenge of adiabaticity becomes more pronounced for ordered state preparation, as it requires passing through the gap minimum. To capture this effect, we introduce an additional length scale $\xi_{\mathrm{ad}}$ which limits long-distance correlations.

\begin{figure}
   \centering
    \includegraphics[width= 1\linewidth]{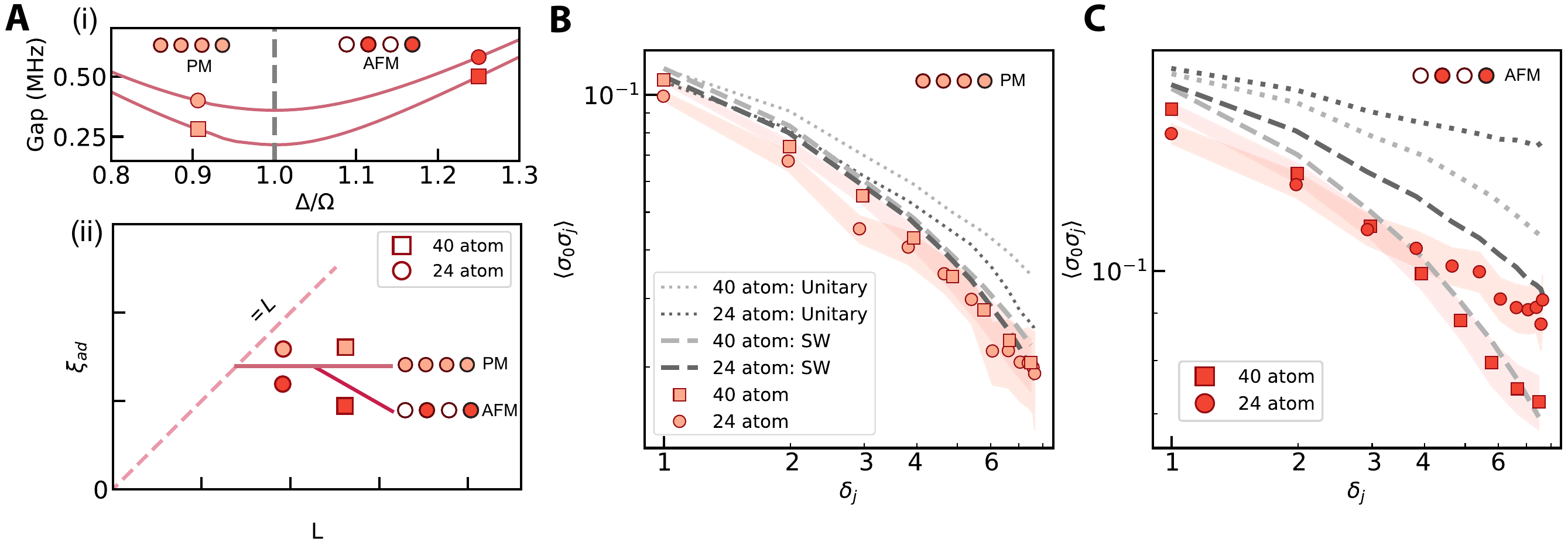}
    \caption{\textbf{The effect of non-adiabaticity}. Circles (24-atom) and squares (40-atom) denote experimental measurements, with the shaded region representing 1-$\sigma$ bootstrap error. Gray dotted and dashed lines indicate unitary, and stochastic wave-function simulation results using experimental parameters. (\textbf{A}) (i) Gap profiles for 24-atom and 40-atom systems. The vertical gray dashed line denotes the location of the gap minimum in the 24-atom system. Markers in the PM (AFM) phase correspond to measurements in B (C). (ii) A schematic showing the dependence of $\xi_{ad}$ on system size using the same LILA ramp for state preparation. (\textbf{B}) and (\textbf{C}): $\sigma$ field correlation measurements in the PM and AFM phases. The influence of decoherence in addition to non-adiabaticity is shown by a comparison between unitary and stochastic wavefunction simulation in (C), indicating that the decoherence affects both 24-atom and 40-atom systems similarly via $\xi_d$. }
    \label{fig:non-adiabaticity}
\end{figure}

Comparing 24-atom and 40-atom correlation measurements with a fixed-time LILA ramp in the disordered (PM) and AFM phases illustrates the above insights. In the PM phase, both the 24-atom and 40-atom correlators exhibit similar behavior (Figure~\ref{fig:non-adiabaticity}B). However, in the AFM phase, the 40-atom correlator decays more rapidly than the 24-atom correlator, resulting in a noticeable deviation at greater distances. This trend can be captured by unitary numerical simulations using LILA ramps, indicating that $\xi_{ad}$ hinders order parameter growth in the 40-atom system and leads to a faster correlator decay in the AFM phase. An exponential fit to the 40-atom correlator data in the AFM phase yields a smaller value (9.8(12)) compared to the decoherence length scale $\xi_d$, confirming the contribution from $\xi_{ad}$.

\subsubsection*{Atom motion}
\begin{figure}
   \centering
    \includegraphics[width= 0.95\linewidth]{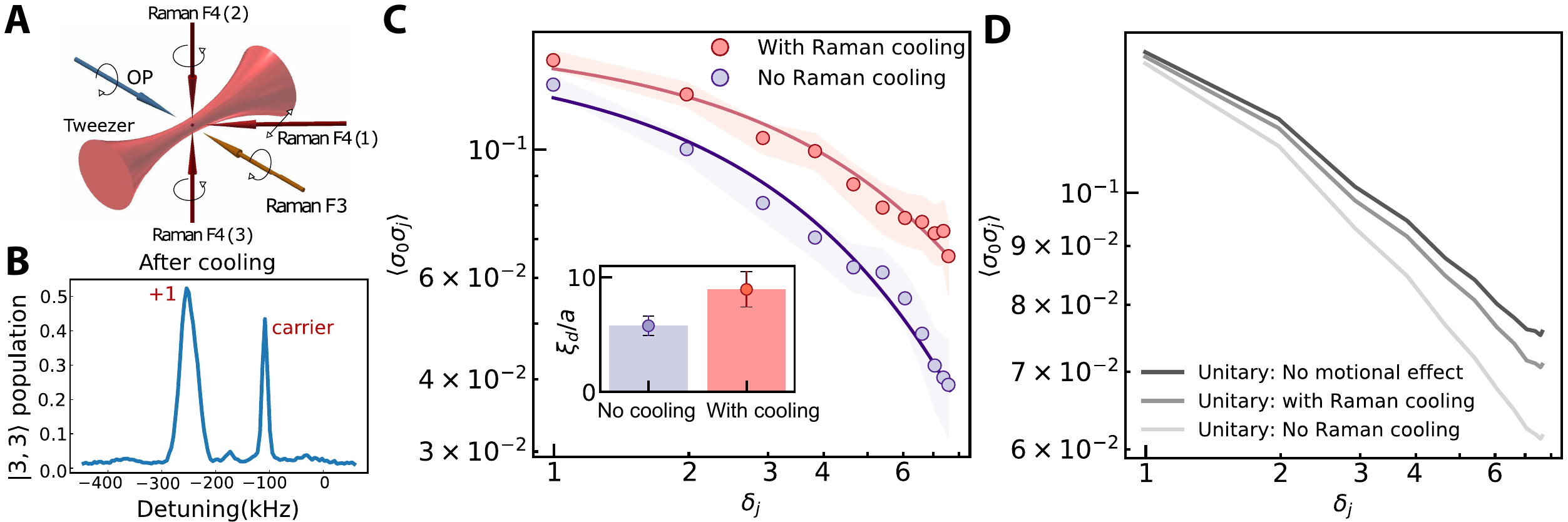}
    \caption{\textbf{Atom motion}. Circles denote experimental measurements, with the shaded region representing 1-$\sigma$ bootstrap error. (\textbf{A}) Raman laser beams configuration, allowing motional state cooling in all three dimensions. (\textbf{B}) Radial sideband thermometry after RSC averaged over the atom array. (\textbf{C}) Measured $\sigma$ field correlations with and without RSC. Inset: Fitted $\xi_d/a$. (\textbf{D}) Simulated $\sigma$ field correlations at the critical point with and without accounting for atom motion using experimental parameters. This motional effect is considered by performing Monte-Carlo simulation. We sample the initial position and momentum distribution and then evolve the positions of the atoms with zero acceleration. Atom temperature is calibrated via release-recapture experiment to be about 5 $\mu$K without Raman cooling. This simulation shows that our critical correlation is minimally affected by atom motion when RSC is applied. }
    \label{fig:RSC}
\end{figure}

\addcontentsline{toc}{subsection}{Atom motion}
Motion of the atoms during the interaction time can cause dynamical variation of interatomic distances, which in turn modifies the distance-dependent interaction strength between atoms. We mitigate motional effects by means of Raman sideband cooling (RSC) into the tweezers. We experimentally evaluate the effect of RSC by comparing the measured  $\langle \sigma_0 \sigma_j \rangle$ correlator in the AFM phase with and without performing RSC (Figure~\ref{fig:RSC} C). Exponential fits show an increase in $\xi_\mathrm{d}$ when RSC is performed, confirming its benefits. 

\subsubsection*{Interaction inhomogeneity}
\addcontentsline{toc}{subsection}{Interaction inhomogeneity}
Optical aberrations in our tweezer system can cause additional non-uniformities in the atomic spacings, leading to a static interaction inhomogeneity. We measure this by calibrating pairwise interaction energies, employing the method from \cite{levine2021}. We rearrange atoms into pairs that are within the blockade radius, and use a pulse to excite them to the symmetric excited state $|W\rangle = \frac{1}{\sqrt{2}}(|gr\rangle + |rg\rangle)$. A second pulse is applied with variable detuning. When the detuning $\Delta = V$, we observe a resonance corresponding to double-excitation into $|rr\rangle$, allowing us to extract the value of $V$. Using this approach, we calibrate the static interaction inhomogeneity (standard deviation/mean) to be 26\% (24-atom ring), 20\% (40-atom ring), and 14\% (rectangular array). We numerically assess the impact of static interaction inhomogeneity and determine that it influences the location of the gap minimum, and does not compromise the fidelity of our critical state preparation. We thus take this inhomogeneity into account in our gap profile calculation and compare it to the experimentally determined critical point. 

\subsection*{Data Processing and Analysis}\label{sec:post-select}
\addcontentsline{toc}{section}{Data Processing and Analysis}
\subsubsection*{Locating the critical point}
\begin{figure}
   \centering
    \includegraphics[width= 0.7\linewidth]{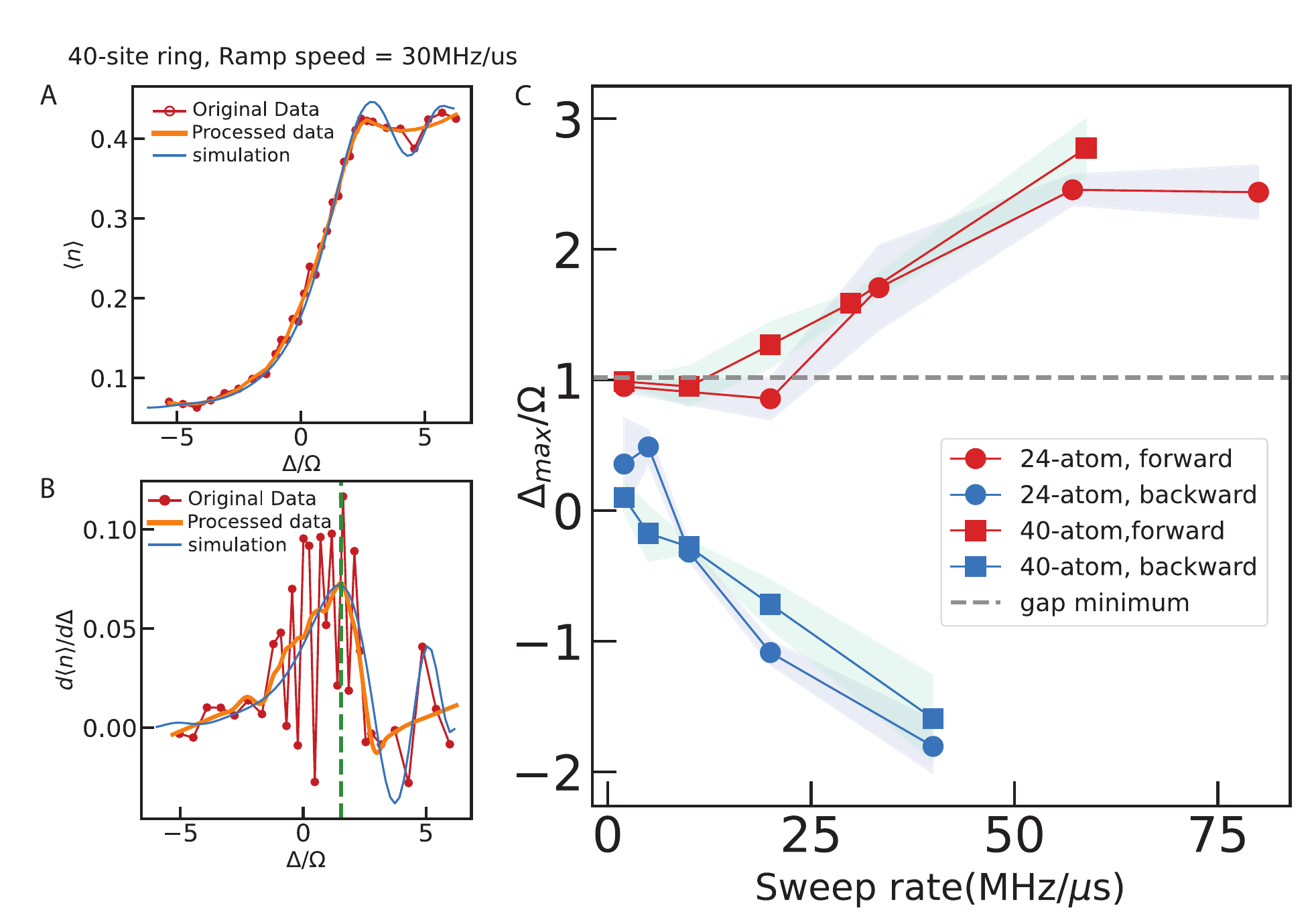}
    \caption{\textbf{Locating critical point for 1D Rydberg array (PBC)}. (\textbf{A}) A typical Rydberg population density at a given ramp speed ($\SI{30}{\mega\hertz/\micro\second}$). Orange curve shows the results after smoothing and interpolation from the experimental measured data (red dots). Blue curve represents the numerical simulation of stochastic evolution of the system considering decoherence and atom loss. (\textbf{B}) A typical susceptibility curve as a function of detuning, further calculated from (A). Red dots shows numerical differentiation to experimentally measured data. The asymmetric shape of the peak and the fast oscillation generated from the numerical differentiation inhibit the accurate extraction of the peak location. The orange curve shows the results after numerical differentiation and another round of data smoothing from the smoothed data in (A). The blue curve shows the numerical differentiation for the numerical simulation, which matches the peak location $\Delta_{\mathrm{max}}$ (dashed green) of the orange curve and validates the extraction scheme. (\textbf{C}) Extracted susceptibility peak location $\Delta_{\mathrm{max}}$ as a function of linear sweep rate. Dots and squares represent the system size of 24 and 40 respectively. Red and blue represent extracted $\Delta_{\mathrm{max}}$ for the forward scan (starting in the paramagnetic phase and linearly ramping up the detuning) and the backward scan (starting in the antiferromagnetic state and linearly ramping down the detuning). Shaded regions represent 1-$\sigma$ bootstrap errors. Ideally, for ramps that are adiabatic, one expects the results from both directions to converge to the critical point (which is numerically calculated from the location of the energy gap minimum for a finite system). The main limitations of the backward scan are decoherence and atom loss during the AFM state preparation.}
    \label{fig:locating_crit_point}
\end{figure}

\addcontentsline{toc}{subsection}{Locating the critical point}
We locate the critical point by experimentally determining the peak of susceptibility $\chi = \frac{d \langle n\rangle}{d \Delta}$ as a function of detuning~\cite{Ebadi2021, Damski20}. However, extracting this peak is challenging due to its asymmetric shape, which hinders accurate extraction using commonly employed polynomial or Gaussian fit procedures. Furthermore, the location of the peak is susceptible to errors induced by the numerical derivative process. We develop a scheme that does not assume a symmetric peak and is more robust to noise, which includes data smoothing and interpolation. The parameters of the data smoothing and interpolation steps are chosen such that it works effectively even when reducing the number of data points of the numerical results (about 300) to match the typical experimental condition of about 50 data points. The key steps of our scheme to extract a $\chi_c = \left(\frac{d \langle n\rangle}{d \Delta}\right)_\mathrm{max}$ are outlined as follows:
\begin{enumerate}
    \item Smooth the data for $\langle n\rangle$ as a function of $\Delta$. We choose a relatively small smoothing window size in the Savitzky-Golay filtering algorithm to identify and retain local features.
    \item Interpolate the smoothed data (orange curve in Fig. \ref{fig:locating_crit_point}A). The interpolation step increases the number of data points and helps stabilize the process of taking the numerical derivative. 
    \item Take the numerical derivative with the central difference method to obtain the curve of $\chi$ as a function of $\Delta$.
    \item Smooth the obtained $\chi$ as a function of $\Delta$ and identify the peak of $\chi$ and corresponding $\Delta_{\mathrm{max}}$, as shown in Fig. \ref{fig:locating_crit_point}B. In this step we choose a relatively large smoothing window size in the Savitzky-Golay filtering algorithm so that we are less vulnerable to oscillations originating from experimental noise. 
\end{enumerate}

This multi-step scheme allows us to accurately determine location of the maximum susceptibility $\Delta_{\mathrm{max}}$. We apply the scheme to the data obtained at various ramp speeds to further confirm the slowest ramp is in the adiabatic region, allowing us to determine the critical detuning $\Delta_c$. Furthermore, in faster ramps, the Rydberg population lags behind the expectation from the ground state, which results in a susceptibility that peaks beyond the true critical point.

Thus far, we have focused on sweeping the detuning from the paramagnetic (PM) to the antiferromagnetic (AFM) phase. As a method to locate the critical point, Ref. \cite{Damski20} suggests using a ramp from the AFM to the PM phase, where an adiabatic ramp would also result in a peak susceptibility at the critical point. Faster ramps would once again result in a lagging Rydberg population, and thus a susceptibility which peaks below the true critical point. We perform the same measurement in a backward ramp scenario, where we adiabatically prepare the AFM state and then linearly sweep down the detuning at different rates. We observe the inverse trend of $\Delta_{\mathrm{max}}$ as a function of sweep rate compared to the forward ramp,  consistent with the prediction described in \cite{Damski20} under the presence of imperfect AFM state preparation (Fig. \ref{fig:locating_crit_point}C). 

\subsubsection*{Postselection of Data}
\addcontentsline{toc}{subsection}{Postselection of Data}
In our experiment, we post-select all our data on successful rearrangement. However, there is approximately 3\% loss per trap from the fluorescence imaging, and an additional 2\% loss per trap from collisions with the background gas during RSC, resulting in a total 5\% loss per trap during state initialization before performing our Rydberg experiments. This loss introduces unknown holes into the perfect 1D ring or 2D rectangular array. These defects are hard to distinguish from Rydberg atoms, which are also detected as loss. However, we note that when Rydberg atoms are created in our sequences, they are more likely to be created at edges or where there are fewer neighboring atoms. Thus, holes due to atom loss are more likely to have neighboring holes that result from an atom being promoted to a Rydberg state. In our data analysis, this creates an atom configuration which violates the so-called Rydberg blockade; namely, there appear to be two or more Rydberg excitations within a blockade radius. To reduce the impact of loss on our data, we postselect on snapshots that do not contain any blockade violations.

\begin{figure}
   \centering
    \includegraphics[width= 1\linewidth]{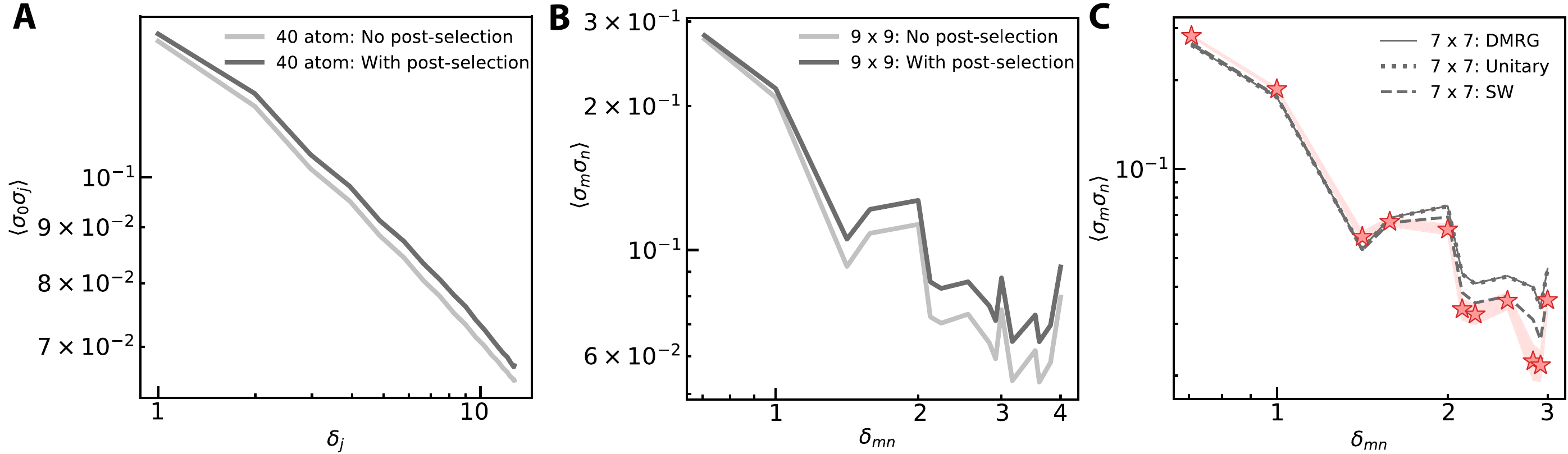}
    \caption{\textbf{Data post-selection}. Stars denote experimental measurements, with the shaded region representing 1-$\sigma$ bootstrap error. Gray solid, dotted, and dashed lines indicate ground (DMRG), unitary, and stochastic wave-function simulation results using experimental parameters. (\textbf{A}, \textbf{B}) Ground state simulations with and without blockade violation post-selection for 40-atom ring (A) and 9 $\times$ 9 rectangular array (B). (\textbf{C}) Raw measurement of the critical $\sigma$-field correlation with loss-inclusive simulations in the 7 $\times$ 7 rectangular array.}
    \label{fig:Data_postSelection}
\end{figure}

To justify the above approach, we confirm numerically that postselection of blockade-violating snapshots does not bias our results. We first calculate the percentage of snapshots in the critical state to contain blockade violations to be 8\% (24-atom ring), 10\% (40-atom ring), 35\% (7 × 7 array), and 52\% (9 × 9 array). Although the percentage is high in 2D, the effect is smaller compared to 1D where holes change the boundary conditions. We calculate the $\sigma$ field correlation of both the entire wavefunction and the portion that contains no blockade-violations. Comparing these two indicates only a minor deviation, suggesting that blockade-violating snapshots do not contribute significantly to this particular observable (Figure~\ref{fig:Data_postSelection} A, B). Furthermore, we can account for atom loss in simulations by averaging over arrays with defects. These simulations agree well our raw data without blockade-violation post-selection, verifying our understanding of the impact of holes on our critical state (Figure~\ref{fig:Data_postSelection} C). All data reported in the main text has been post-selected to exclude blockade violations, with the exception of data used for locating the critical point via linear ramps.

\subsubsection*{Bootstrap Errorbars}
\addcontentsline{toc}{subsection}{Bootstrap Errorbars}
In this paper, we use bootstrap methods~\cite{bootstrap_Ana} to estimate errorbars. From $ N $ snapshots of a particular experiment, we sample $ N $ times with replacement. We then proceed to calculate all observables, such as $\sigma$ field correlators, and perform all required fits. We repeat this procedure for $ B $ (typically $\geq$ 1000) bootstrap samples, and enumerate the calculated values. The reported value for the quantity is then taken to be the mean of the bootstrapped values with an errorbar given by the standard deviation. If the distribution of the bootstrapped values is asymmetric, we report the median value as our best estimate of the quantity with an errorbar determined by the 15.8$th$ and 84$th$ percentile of the distribution. This procedure results in the asymmetric errorbar in $\xi_d/a$ reported in the main text, extracted from the power-law exponential fits (Fig. 2E and Fig. 3A,B). For all quantities reported in this work, including $\Delta_\sigma^{{}^{{}_{\text{1D}}}}$, $\Delta_\sigma^{{}^{{}_{\text{2D}}}}$ and $\xi_d/a$, we incorporate both the bootstrap and fitting errorbars. 

\subsection*{Boundary phase transitions in 2D}
\begin{figure}
   \centering
    \includegraphics[width= 1\linewidth]{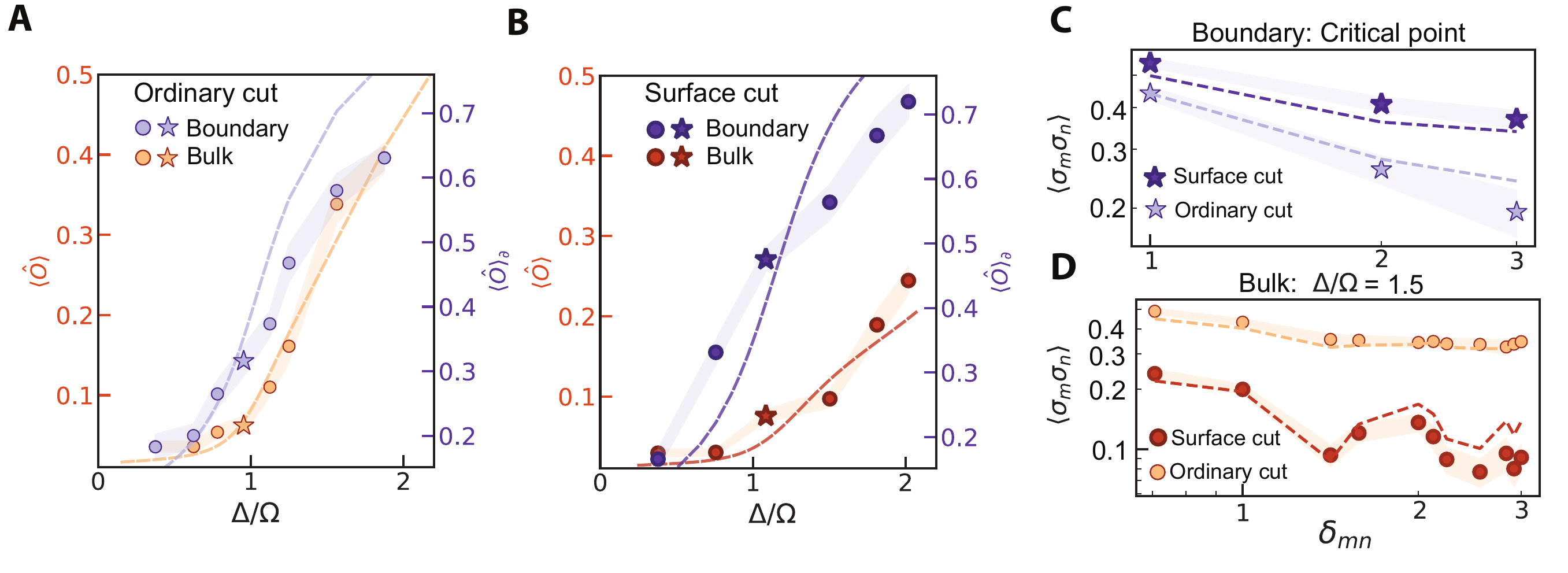}
    \caption{{\bf 2D ordinary vs surface transitions.} Dashed lines represent ground-state numerical simulations using experimental parameters. (\textbf{A}, \textbf{B}) The order parameter, analyzed separately for the bulk ($\langle \hat{O} \rangle$) and boundary ($\langle \hat{O} \rangle_\partial$), is shown across the phase transition for the ordinary cut (A) and the surface cut (B). (\textbf{C}, \textbf{D}) Measured spatial correlations of the $\sigma$ field along the boundary (C) at the critical point for both cuts, and within the bulk (D) at $\Delta/\Omega = 1.5$.}
    \label{fig:SF_Org}
\end{figure}

\addcontentsline{toc}{section}{Boundary phase transitions in 2D}
In the main text, we detailed the experimental observation of two boundary universality classes: the surface class, where the boundary orders independently of the bulk, and the ordinary class, where the bulk and boundary order simultaneously~\cite{AJBray_1977, Burkhardt94, Gliozzi2015}. We noted that the decay of the two-point correlator $\sigma_{i,j}$ (Eqn. \ref{eq:sigma2D}) was slower for the surface transition than the ordinary transition. This is qualitatively in agreement with theoretical expectations. As $\sigma_{i,j}$ is odd under the $\mathbb{Z}_2$ Ising symmetry of the transition, we should thus expect that asymptotically its decay is governed by the scaling dimension the most relevant odd CFT field on the boundary. For the surface transition, this is just the scaling dimension of the usual (1+1)d Ising universality class spin field, $\Delta_\sigma^{{}^{{}_{\text{1D}}}}=1/8$. By contrast, for the ordinary transition the scaling dimension for the most relevant odd boundary field has instead been calculated via conformal bootstrap to be much greater: $\Delta_{\partial_z \hat{\sigma}}=1.276(2)$~\cite{Gliozzi2015}. Nevertheless, it is still instructive to quantitatively compare our experimental results with numerical predictions under the same conditions. To this end, we perform ground state simulations in 2D using DMRG across both the ordinary and surface cuts that are explored in the experiments; in order to best compare with the data, we include a small, experimentally-measured hole-fraction and implement the same type of post-selection. As shown in Fig.~\ref{fig:SF_Org}, our numerical simulations (dashed lines) agree quite well with the experimental data, reproducing the same qualitative differences at the surface versus ordinary cut. 

\subsection*{Numerical studies}
\addcontentsline{toc}{section}{Numerical studies}
We utilize matrix product state (MPS) methods for numerical simulations throughout our work, using the finite-size density matrix renormalization group (DMRG) algorithm for ground state studies and the two-site time-dependent variational principle (TDVP) method for time dynamics~\cite{mps18, SCHOLLWOCK201196, Haegeman16}.

\subsubsection*{Stochastic wavefunction numerics} \label{sect:noise_op}
\addcontentsline{toc}{subsection}{Stochastic wavefunction numerics}
To capture the openness of our quantum system, we perform simulations of its time dynamics incorporating the two most relevant sources of single-body decoherence in our experiment, intermediate-state scattering and Rydberg state decay. This is generically described by the Lindblad master equation,
\begin{align}
\frac{d \rho}{d t} = -\frac{i}{\hbar} \left[ H, \rho \right] + \sum_j \left( c_j \rho c_j^\dagger - \frac{1}{2} \left\{ c_j^\dagger c_j \right\} \right),
\end{align}
where $\rho$ is the density matrix, $H$ is the system Hamiltonian, and the $c_j$ are the jump operators describing the decoherence channels.

We model the Rydberg decay process using the jump operators $c_i^{\mathrm{decay}} = \sqrt{\gamma_{\mathrm{decay}}} \left| g \right>_i\prescript{}{i}{\left< r \right|}$ for each atom $i$, where $1/\gamma_{\mathrm{decay}}=71.44\mskip\thinmuskip\mu s$ is the lifetime of the Rydberg state. To model intermediate-state scattering, we first quantify the two-photon excitation process as a driving field $\Omega_{\mathrm{blue}}=2\pi\times 80.04\mskip\thinmuskip \mathrm{MHz}$ from the ground state $\left| g \right>$ to an intermediate state $\left| e \right>$ with a detuning $\delta=2\pi\times 1.058\mskip\thinmuskip \mathrm{GHz}$, and a second driving field $\Omega_{\mathrm{IR}}=2\pi\times 42.3\mskip\thinmuskip \mathrm{MHz}$ from $\left| e \right>$ to the Rydberg state $\left| r \right>$. The intermediate state decays at a rate $\gamma_e=2\pi\times 1.23\mskip\thinmuskip \mathrm{MHz}$, which we then model with the jump operators $\sqrt{\gamma_e} \left| g \right>_i \prescript{}{i}{\left< e \right|}$ for each atom $i$. Next, we perform adiabatic elimination on the intermediate state $\left| e \right>$, leading to an effective Lindbladian with a two-photon Rabi frequency $\Omega = \frac{\Omega_{\mathrm{blue}} \Omega_{\mathrm{IR}}}{2 \delta}=2\pi\times 1.6\mskip\thinmuskip \mathrm{MHz}$ and effective jump operators $c_i^{\mathrm{scatt}} = \sqrt{\frac{\gamma_e}{4 \delta^2}} \left( \Omega_{\mathrm{blue}} \left| g \right>_i \prescript{}{i}{\left< g \right|} + \Omega_{\mathrm{IR}} \left| g \right>_i \prescript{}{i}{\left< r \right|} \right)$ for each atom $i$. The intermediate-state scattering process can be summarized using an effective rate $\gamma_{\mathrm{scatt}} = \frac{\gamma_e}{4 \delta^2} \left( \Omega_{\mathrm{blue}}^2 + \Omega_{\mathrm{IR}}^2 \right) = 1/70.69\mskip\thinmuskip \mu s$, which is of a similar timescale to the Rydberg lifetime. We note that the presence of these two decoherence channels has a significant effect on the dynamics of the system, which occurs within a time period of $7 \mskip\thinmuskip \mu s$.

We implement open-system time evolution using the stochastic wavefunction framework, where we stochastically generate quantum trajectories of the system to produce an ensemble of pure states that can be averaged over to recover the behavior of the density matrix obtained using the Lindblad master equation. Under decoherence, quantum state evolution is modified from the Schr\"{o}dinger equation such that the system evolves under a non-Hermitian effective Hamiltonian $H_{\mathrm{eff}} = H - \frac{i \hbar}{2} \sum_j c_j^\dagger c_j$ and, for every jump operator $c_j$, the wavefunction $\left| \psi \right>$
stochastically ``jumps'' to $\frac{c_j \left| \psi \right>}{\left< \psi \right| c_j^\dagger c_j \left| \psi \right>}$ with probability density $\left< \psi \right| c_j^\dagger c_j \left| \psi \right>$ in time. In our numerics, we average over $O \left( 100 \right)$ such trajectories per simulation.

\subsubsection*{Additional numerical results at larger system sizes}
\addcontentsline{toc}{subsection}{Additional numerical results at larger system sizes}
To understand the validity and scalability of our results, we run numerics on larger system sizes in 1D for both the ground state (up to $L=200$) and time dynamics with simulated decoherence (up to $L=100$), and confirm that they remain consistent with our main results.
In particular, we study the impact of finite size on three key features: (i) the location of the quantum critical point, (ii) the extracted critical exponent of the $\sigma$ field, and (iii) the decoherence-induced length-scale. 

\begin{figure}
   \centering
    \includegraphics[width=\linewidth]{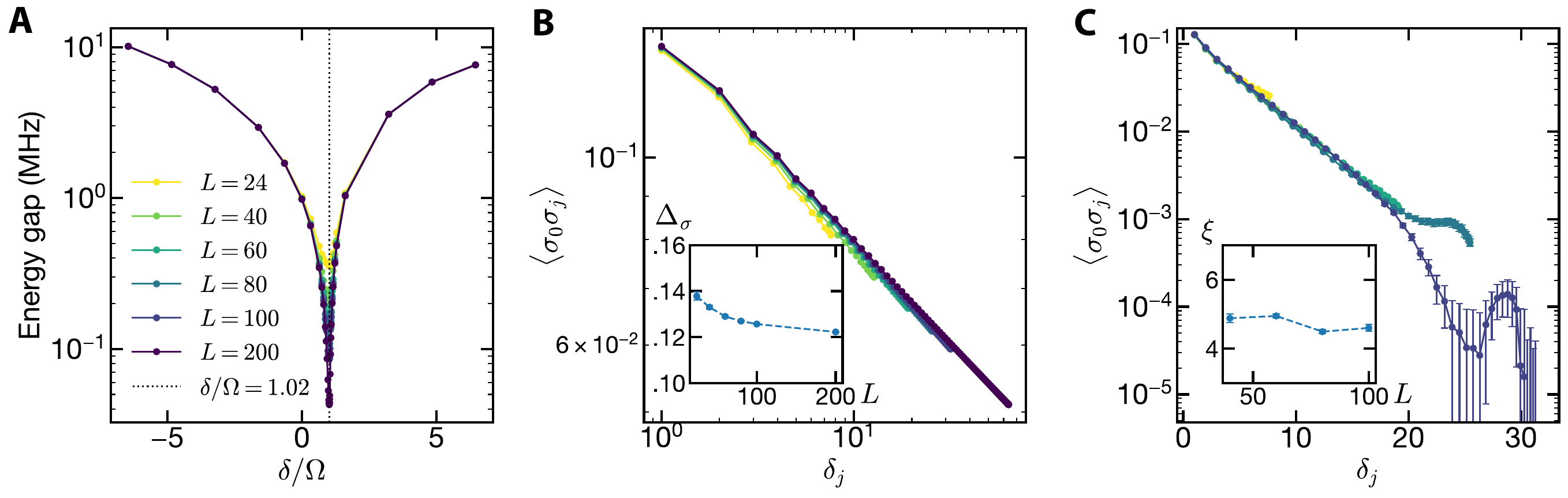}
    \caption{{\bf Numerical results for larger system sizes in 1D.} (\textbf{A}) Energy gap as a function of $\delta/\Omega$ for system sizes up to $L=200$. The gap minimum and extracted critical point remain consistent at $\delta/\Omega=1.02$. (\textbf{B}) Ground state $\sigma$ field correlations as a function of chord distance at the extracted critical point for system sizes of up to $L=200$, showing consistent power-law behavior with similar power-law exponents for different sizes (inset: extracted critical exponent vs system size). (\textbf{C}) $\sigma$ field correlations on a semilog scale for a $t=15 \mu \mathrm{s}$ dynamical LILA ramp with decoherence effects included, using energy gaps determined by part (A), showing consistent behavior across all system sizes (inset: extracted exponential length scale vs system size). 
    The error bars represent uncertainty from Monte Carlo sampling in the stochastic wavefunction method.}
    \label{fig:bigL}
\end{figure}

The results are summarized in Fig.~\ref{fig:bigL}.
From ground state simulations, we find that the extracted critical point from the location of the minimum energy gap is consistent across system sizes from $L=24$ to $L=200$ (see Fig.~\ref{fig:bigL}).
Moreover, as shown in Fig.~\ref{fig:bigL}B, the decay of the 2-point $\sigma$ field correlator as a function of chord distance, $\delta_j$, exhibits an exponent (slope on the log-log plot) that quickly converges to the theoretical expectation as a function of system size [inset, Fig.~\ref{fig:bigL}B]; for $L \ge 40$, the extracted critical exponent only changes by $<10\%$ up to $L=200$.

We now turn to large-scale master equation simulations in order to understand the role of finite size on the emergent, decoherence-induced length scale.
In particular, we directly simulate the adiabatic preparation protocol using  the same  ``local linear adiabatic'' ramp profile as in the experiment and in the presence of the noise operators described in section \ref{sect:noise_op}.
In simulating system sizes up to $L=100$, we note that one must utilize longer adiabatic ramp times because the finite-size gap at the critical point decreases as $\sim1/L$; to this end, for all of the subsequent simulations, we utilize a  ramp time of $t=15 \mu \mathrm{s}$.
At this timescale, the decoherence-induced exponential behavior dominates over the  slower critical power law, as shown in Fig.~\ref{fig:bigL}C where the correlations of the $\sigma$ field are nearly linear on a semilog scale, exhibiting deviations only at the shortest distances. 
These correlations are remarkably consistent across all system sizes, both in their absolute values and in the exponential slope, supporting our hypothesis that decoherence yields a single emergent, cutoff length-scale independent of system size.
We also quantitatively fit the exponential length scale (Fig.~\ref{fig:bigL}C inset) for each system size using a combined power law times exponential functional form: as expected, the resulting values of $\xi$ are consistent across all system sizes.
%
%
We also note that for the largest system sizes, oscillations appear at large distances owing to effects from non-adiabaticity; by taking even longer ramps, one could in principle suppress these effects. 

\begin{figure}
   \centering
    \includegraphics[width=0.4\linewidth]{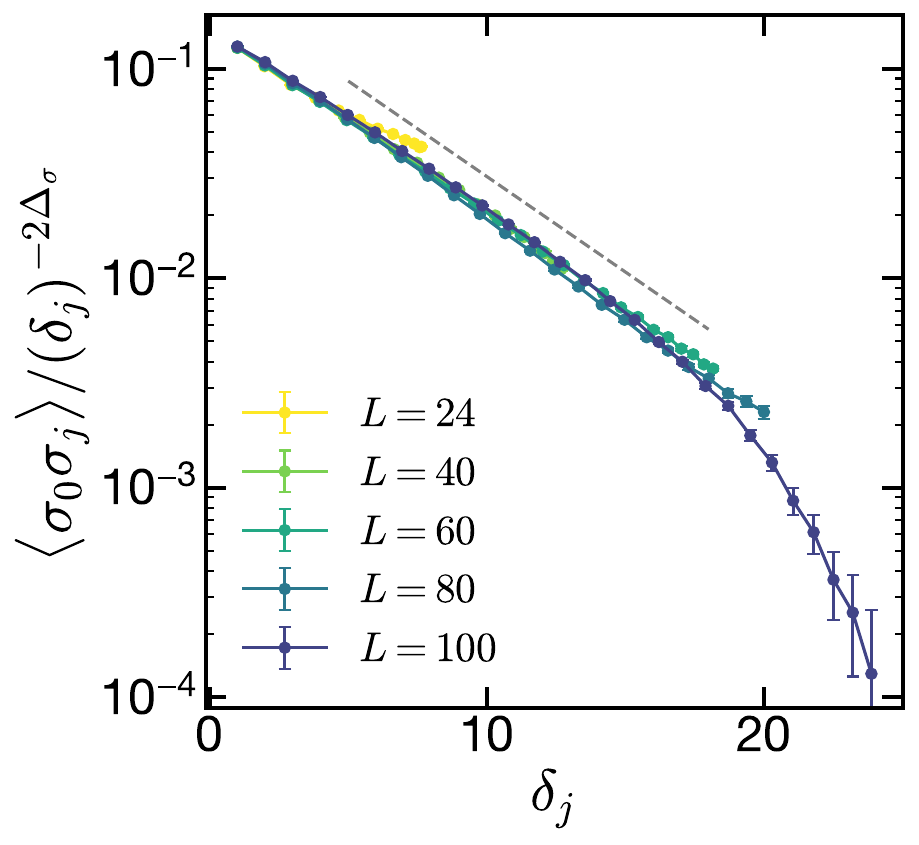}
    \caption{Semilog plot of two-point $\sigma$ field correlations in master equation simulations (same as Fig.~\ref{fig:bigL}C) divided by theoretical power law scaling $\Delta_\sigma=1/8$, as a function of distance. The oscillations for $L=80$ and $L=100$ are truncated away.}
    \label{fig:exp}
\end{figure}

We can also use these additional numerics to further validate our fitting procedures in the main text.
In particular, we are able to verify that our functional form for fitting, that of a power law times an exponential, is the qualitatively correct form to capture the system's behavior in the presence of decoherence even for larger system sizes.
As previously mentioned, the critical exponent is difficult to quantitatively extract in longer ramps due to the dominance of the exponential decay, but Fig.~\ref{fig:bigL}C shows a noticeable curve at low $\delta_j$ on a semilog plot indicating a deviation from purely exponential behavior.
If the functional form is a power law times an exponential, we expect this initial curve to be flattened by dividing by the power law, becoming fully linear.
Thus, we plot the $\sigma$ field correlations normalized by the critical power-law decay $\left( \delta_j \right)^{-2 \Delta_\sigma}$ (Fig.~\ref{fig:exp}), using the theoretical value of $\Delta_\sigma=1/8$, in Fig.~\ref{fig:exp}.
Indeed, the plots are linear for every system size studied (except $L=100$, where the nonlinearity only begins significantly after we expect non-adiabatic effects at around distance 10).
Thus, our numerical results for $\sigma$ field correlations are fully consistent with power law times exponential behavior, even to larger system sizes, validating our functional form and, by extension, our extraction of the critical exponent $\Delta_\sigma$, in the main text.

\subsubsection*{Theoretical motivation for decohered correlations}
\addcontentsline{toc}{subsection}{Theoretical motivation for decohered correlations}

The preceding numerics establish that the critical state prepared in the presence of decoherence shows correlations that decay as a power law multiplied by an exponential. 
Here, we provide some intuition for why such a profile can arise.
A broad theoretical understanding of quantum criticality in the presence of decoherence is still being actively developed.
A simple and likely scenario is when decoherence (or unitary noise) acts similarly to a finite-temperature bath. 
This has previously been shown to be the case in a variety of open and noisy quantum systems~\cite{Mitra:2006,Diehl:2008,Maghrebi:2016,Paz:2021b,Dupont:2022}. 
Then, one can gain intuition from known results of quantum criticality at finite temperatures. 
For example, in a 1+1D conformal field theory at finite temperature $T$~\cite{Cardy:1984}, the exact form of the correlation function  is
\begin{equation} 
    \langle \sigma_0\sigma_j\rangle \propto \left(\frac{T}{\sinh(\pi T j)}\right)^{-2\Delta_\sigma^{{}^{{}_{\text{1D}}}}}.
\end{equation}
This has two key features that are generic to weakly-perturbed critical points: (i) power-law decay at short distances, featuring the true critical exponents, and (ii) a crossover to an exponential decay at long distances.
Such behavior is seen within so-called quantum critical fans, i.e. extended regions in phase space where the correlations are still governed by the zero-temperature quantum critical point, but limited by a length-scale imposed by the thermal fluctuations~\cite{Sachdev:1997}. 
The experimental and numerical evidence suggests that our prepared state exists within some general quantum critical fan, with the total decoherence (i.e. integrated over the finite-time dynamics) acting akin to an effective temperature.
This does not dictate the precise form of the correlation decay profile, but the simplest function with the right features is $C(r) \sim r^{-2\Delta} e^{-r/\xi}$.
This simple function seems to fit the data well [Fig.~\ref{fig:bigL},~\ref{fig:exp} and preceding discussion]; we note that a direct multiplication by an exponential decay also appeared in Ref.~\cite{Dupont:2022}, which similarly studied the preparation of the Ising critical point in the presence of noise.

\subsection*{Energy Field}
\addcontentsline{toc}{section}{Energy Field}
In the main text, we focus on the primary field $\sigma$, corresponding to the spin field of the Ising CFT. The other primary field $\epsilon$, known as the energy field, can be similarly represented, at the leading order in 1D, by a microscopic lattice operator \cite{Slagle2021}
\begin{equation}
\epsilon_{i+1/2} = (n_i + n_{i+1}) - \langle n \rangle.
\end{equation} 
The two-point correlator $\langle \epsilon_{1/2}\epsilon_{j+1/2}\rangle$ decays as a power law
\begin{equation}
\langle \epsilon_{1/2}\epsilon_{j+1/2}\rangle \propto \delta_j^{-2\Delta_\epsilon^{{}^{{}_{\text{1D}}}}},
\end{equation} 
where $\Delta_\epsilon^{{}^{{}_{\text{1D}}}} = 1$, predicted by the (1+1)d Ising CFT.

Similar to the case of the $\sigma$ field, our adiabatic state preparation protocol provides direct access to the $\epsilon$ field. As shown in Fig. \ref{fig:eps2DDecohere}A, we observe a pronounced decay of $\langle \epsilon_{1/2}\epsilon_{j+1/2}\rangle$. However, two main challenges prevent us from determining $\Delta_\epsilon^{{}^{{}_{\text{1D}}}}$ experimentally. First, this large scaling dimension $\Delta_\epsilon^{{}^{{}_{\text{1D}}}}$ leads to a more rapid decay of the $\langle \epsilon_{1/2}\epsilon_{j+1/2}\rangle$ compared to $\langle \sigma_{0}\sigma_{j}\rangle$. The measured two-point correlator signal quickly drops to the noise floor. Increasing the system size does not help in this scenario (Fig. \ref{fig:eps2DDecohere}A). Second, in contrast to the 2D case reported in the main text ($\Delta_\sigma^{{}^{{}_{\text{2D}}}} \approx 0.5$), the 1D geometry has limited access to the spatial separations between the primary fields. Therefore, insufficient measurement points are available to fit a power law before the signal descends below the noise floor. In addition to 1D, the (2+1)d Ising CFT predicts $\Delta_\epsilon^{{}^{{}_{\text{2D}}}}\approx1.4$. This larger scaling dimension makes it challenging to observe the power law decay even in the 2D rectangular array where more spatial separations between the primary fields are accessible. 

\subsection*{2D boundary decoherence}
\begin{figure}
   \centering
    \includegraphics[width= 0.8\linewidth]{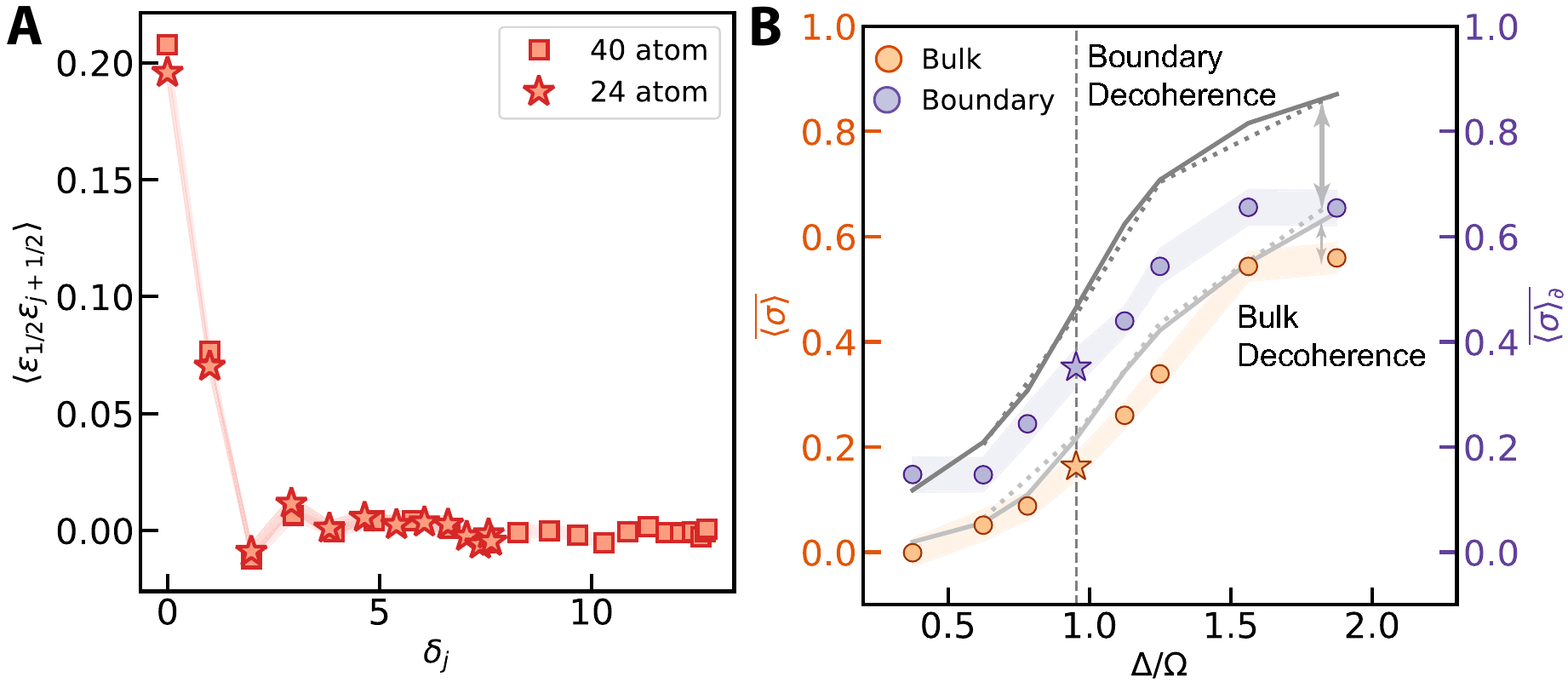}
    \caption{(\textbf{A}) Two-point correlations of the energy field $\epsilon$ at the critical point for both 24-atom and 40-atom systems. (\textbf{B}) Averaged one-point function analyzed independently for the bulk ($\overline{\langle \sigma \rangle}$) and the boundary ($\overline{\langle \sigma \rangle_{\partial}}$) across the phase transition. }
    \label{fig:eps2DDecohere}
\end{figure}

\addcontentsline{toc}{section}{2D boundary decoherence}
The odd-length, open boundary condition in 2D explored in this work explicitly breaks the $\mathbb{Z}_2$ symmetry and leads to a unique ground state in the ordered phase. Because of this, the one-point function $\langle \sigma \rangle$ is non-vanishing. It transitions from 0 in the fully disordered phase to 1 in the perfect checkerboard phase (panel (ii) in Fig. 4B). We measure ${\langle \sigma \rangle}$ across the critical point (Fig. \ref{fig:eps2DDecohere}B), and separately average over bonds either within the bulk ($\overline{\langle \sigma \rangle}$) or along the boundary ($\overline{\langle \sigma \rangle_{\partial}}$). A comparison between our measurements and ground state expectation (solid gray line in Fig. \ref{fig:eps2DDecohere}B) reveals discrepancies. In particular, $\overline{\langle \sigma \rangle_{\partial}}$ deviates more compared to $\overline{\langle \sigma \rangle}$. Meanwhile, a time-dependent simulation using LILA ramp that accounts for non-adiabatic errors yields a result that aligns well with the ground state expectation (dotted gray line in Fig. \ref{fig:eps2DDecohere}B). This indicates that decoherence is the primary factor and has a more pronounced effect along the boundary. 

This observation stems from the presence of strong corner effects. Atoms located at the corners have fewer neighbors, making them more readily promoted to Rydberg states. Once promoted, these atoms act as external symmetry-breaking fields that drive the growth of the one-point function and enhance the boundary symmetry-breaking in our odd-length geometry. This enhanced symmetry-breaking along the boundary is additionally supported by the greater two-point correlator along the boundary in comparison to that within the bulk (Fig. 4E). These enhanced correlations are themselves more prone to the decoherence-induced length scale $\xi_d$.
\clearpage

\end{document}